\documentclass[12pt]{article}

\pdfoutput=1 

\usepackage{natbib}
\usepackage{hyperref}
\usepackage{rotating}
\usepackage{graphicx}
\usepackage{subfigure}

\usepackage{float}
\usepackage{amsmath} 
\usepackage{amssymb}
\usepackage{amsthm}
\usepackage{mhequ}

\usepackage{color}
\newcommand{\xbeta}{ x_i \beta}
\newcommand{\xtheta}{ x_i \theta}

\newtheorem{remark}{Remark}

\newcommand{\be}{\begin{equs}}
\newcommand{\ee}{\end{equs}}
\newcommand{\bb}[1]{\mathbb{#1}}
\newcommand{\mc}[1]{\mathcal{#1}}
\newcommand{\Binom}{\text{Binomial}}
\newcommand{\No}{\text{No}}
\newcommand{\PG}{\text{PG}}
\newcommand{\IG}{\text{Inverse-Gamma}}

\newcommand{\Bern}{\text{Bernoulli}}
\newcommand{\Poi}{\text{Poisson}}
\newcommand{\NB}{\text{NB}}
\newcommand{\cov}{\text{cov}}
\newcommand{\var}{\text{var}}
\newcommand{\diag}{\text{diag}}

\newcommand{\bigO}{\mc O}
\begin{document}

%\title{Calibrated Data Augmentation for Scalable \\ Markov Chain Monte Carlo}
\title{{Scaling up Data Augmentation MCMC via Calibration}}

\author{ Leo L. Duan\thanks{Department of Statistical Science, Duke University, Durham, NC, email: leo.duan@duke.edu},
     James E. Johndrow\thanks{Department of Statistics, Stanford University, Stanford, CA, email: johndrow@stanford.edu},
     David B. Dunson\thanks{Department of Statistical Science, Duke University, Durham, NC, email: dunson@duke.edu}
     }

\date{}

% \editor{}
\maketitle

{\bf Abstract:} 
{There has been considerable interest in making Bayesian inference more scalable.}
{In big data settings},  most literature focuses on reducing the computing
time per iteration, with less focused on reducing the number of iterations
needed in Markov chain Monte Carlo (MCMC). This article focuses on data
augmentation MCMC (DA-MCMC), a widely used technique. {DA-MCMC samples tend
to become highly autocorrelated in large samples, due to a miscalibration problem in which conditional posterior distributions given augmented data are too concentrated.  This makes it necessary to collect very long MCMC
paths to obtain acceptably low MC error.} {To combat this inefficiency, we propose a family of calibrated data augmentation algorithms, which appropriately adjust the variance of conditional posterior distributions. A Metropolis-Hastings step is used to eliminate bias in the stationary distribution of the resulting sampler}. {Compared to existing alternatives, this approach} {can dramatically reduce MC error by reducing autocorrelation and increasing the effective number of DA-MCMC samples per computing time.} The approach is {simple and}  applicable to a broad variety of existing data augmentation algorithms, and we focus on three popular models: probit, logistic and Poisson log-linear.  Dramatic gains in computational efficiency are shown in applications.
\vskip 12pt 

{\noindent  KEY WORDS:   Bayesian Probit; Bayesian Logit; Big $n$; Data Augmentation; Maximal Correlation; Polya-Gamma for Poisson.}

\vfill

\newpage

\pagenumbering{arabic}

\section{Introduction}

With the deluge of data in many modern application areas, there is pressing need for scalable computational algorithms for inference from such data, including uncertainty quantification (UQ).  Somewhat surprisingly, even as the volume of data increases, uncertainty often remains sizable.  Examples in which this phenomenon occurs include financial fraud detection \citep{ngai2011application}, disease mapping \citep{wakefield2007disease} and online click-through tracking \citep{wang2010click}.  Bayesian approaches provide a useful paradigm for quantifying uncertainty in inferences and predictions in these and other settings.

The standard approach to Bayesian posterior computation is Markov chain Monte Carlo (MCMC) and related sampling algorithms. However,  conventional MCMC algorithms often scale poorly in problem size and complexity. Due to its sequential nature, the computational cost of MCMC is the product of two factors: the evaluation cost at each sampling iteration and the total number of iterations needed to obtain an acceptably low Monte Carlo (MC) error. While a substantial literature has developed focusing on decreasing computational cost per iteration (\cite{minsker2014robust,maclaurin2014firefly,
srivastava2015wasp,conrad2015accelerating} among others), {very little has been done
to reduce the} {number of iterations needed to produce a desired MC error in posterior summaries} 
{as the problem size grows.}

{A major concern in applying MCMC algorithms in big data problems is that the level of autocorrelation in the MCMC path may increase with the size of the data.  Markov chains with high autocorrelation 
tend to produce a low {\em effective sample size (ESS)} per unit computational time, {which is informally known as the  {\em slow mixing} problem}.  The ESS is designed to compare the information content in the sampling iterations relative to a gold standard Monte Carlo algorithm that collects independent samples.  If the number of effective samples in 1,000 iterations is only 10, then the MCMC algorithm will need to be run 100 times as long as the gold standard algorithm to obtain the same MC error in posterior summaries.  Such a scenario is not unusual in big data problems, leading MCMC algorithms to face a {\em double burden}, with the time per iteration increasing and it becoming necessary to collect more and more iterations.} 

{This double burden has led many members of the machine learning community to abandon MCMC in favor of more easily scalable alternatives, such as variational approximations. Unfortunately these approaches lack theoretical guarantees and often badly under-estimate posterior uncertainty.  Hence, there has been substantial interest in recent years in designing scalable MCMC algorithms.  The particular focus of this paper is on a popular and broad class of Data Augmentation (DA)-MCMC algorithms.  DA-MCMC algorithms are used routinely in many classes of models, with the algorithms of \cite{albert1993bayesian} for probit models and \cite{polson2013bayesian} for logistic models particularly popular.  Our focus is on improving the performance of such algorithms in big data settings in which issues can arise in terms of both the time per iteration and the mixing.  The former problem can be addressed using any of a broad variety of existing approaches, and we focus here on the slow
mixing problem.}

{ \cite{johndrow2016inefficiency} discovered that popular DA-MCMC algorithms have small effective sample sizes in large data settings involving imbalanced data.  For example, data may be binary with a high proportion of zeros.  A key insight is that the reason for this problem is a discrepancy in the rates at which Gibbs step sizes and the width of the high-probability region of the posterior converge to zero as $n$ increases.  In particular, the conditional posterior given the augmented data may simply be too concentrated relative to the marginal posterior, with this problem amplified as the data sample size increases.  There is a rich literature on methods for accelerating mixing in DA-MCMC algorithms using tricks ranging from reparameterization to parameter-expansion \citep{liu1999parameter,meng1999seeking,papaspiliopoulos2007general}.  However, we find that such approaches fail to address the miscalibration problem and have no impact on the worsening mixing rate with increasing data sample size $n$.}

{The focus of this article is on proposing a general new class of algorithms for addressing the fundamental miscalibration problem that leads to worsening mixing of  DA-MCMC with $n$.  In particular, the key idea underlying our proposed class of {\em calibrated} DA (CDA) algorithms is to introduce auxiliary parameters that change the variance of full conditional distributions for one or more parameters. These auxiliary parameters can adapt with the data sample size $n$ to fundamentally address the key problem causing the worsening mixing with $n$.  In general, the invariant measure of CDA-MCMC does not correspond exactly to the true joint posterior distribution of interest. Instead, we can view CDA-MCMC as representing a computationally more efficient perturbation of the original Markov chain. { The perturbation error can be eliminated using Metropolis-Hastings. Compared to other adaptive Metropolis-Hastings algorithms,
which often require carefully chosen multivariate proposals and complicated adaptation with multiple chains  \citep{tran2016adaptive}, CDA-MCMC only
requires a simple modification to Gibbs sampling steps. We show the auxiliary parameters can be efficiently adapted for each type of data augmentation, via minimizing
the difference between Fisher information of conditional and
marginal distributions.}}

\section{Calibrated Data Augmentation} \label{sec:cda}
Data augmentation Gibbs samplers alternate between sampling  latent data $z$ from their conditional posterior distribution given model parameters $\theta$ and observed data $y$, and sampling parameters $\theta$ given $z$ and $y$; either of these steps can be further broken down into a series of full conditional sampling steps but we focus for simplicity on algorithms of the form: 
\be \label{eq:da}
z \mid \theta, y &\sim \pi(z;\theta,y) \\
\theta \mid z,y &\sim f(\theta;z,y),
\ee
where $f$ belongs to a location-scale family, such as the Gaussian.  Popular data augmentation algorithms are designed so that both of these sampling steps can be conducted easily and efficiently; e.g., sampling the latent data for each subject independently and then drawing $\theta$ simultaneously (or at least in blocks) from a multivariate Gaussian or other standard distribution.  This effectively avoids the need for tuning, which is a major issue for Metropolis-Hastings algorithms, particularly when $\theta$ is high-dimensional.
Data augmentation algorithms are particularly common for generalized linear models (GLMs), with $\bb E(y_i \mid x_i, \theta) = g^{-1}(x_i \theta)$ and a conditionally Gaussian prior distribution chosen for $\theta$. We focus in particular on Poisson log-linear, binomial logistic, and binomial probit as motivating examples.

Consider a Markov kernel $K((\theta,z);\cdot)$ with invariant measure $\Pi$ and update rule of the form \eqref{eq:da}, and a Markov chain $(\theta_t,z_t)$ on a state space $\Theta \times \mc Z$ evolving according to $K$. We will abuse notation in writing $\Pi(d\theta) = \int_{z \in \mc Z} \Pi(d\theta,dz)$. The lag-1 autocorrelation for a function $g : \Theta \to \bb R$ at stationarity can be expressed as the Bayesian fraction of missing information (\cite{papaspiliopoulos2007general}, \cite{rubin2004multiple}, \cite{liu1994fraction})
\be
\gamma_g &= 1- \frac{\bb E[\var(g(\theta) \mid z)]}{\var(g(\theta))}, \label{eq:missinginfo}
\ee
where the integrals in the numerator are with respect to $\Pi(d\theta,dz)$ and in the denominator with respect to $\Pi(d\theta)$. Let 
\be
L_2(\Pi) = \left\{ g : \Theta \to \bb R, \int_{\theta \in \Theta} \{g(\theta)\}^2 \Pi(d\theta) < \infty \right\} 
\ee
be the set of real-valued, $\Pi$ square-integrable functions. The \emph{maximal autocorrelation}
\be
\gamma = \sup_{g \in L^2(\Pi)} \gamma_g = 1- \inf_{g \in L^2(\Pi)} \frac{\bb E[\var(g(\theta) \mid z)]}{\var(g(\theta))}
\ee
is equal to the geometric convergence rate of the data augmentation Gibbs sampler (\cite{liu1994fraction}). For $g(\theta) = \theta_j$ a coordinate projection, the numerator of the last term of \eqref{eq:missinginfo} is, informally, the average squared step size for the augmentation algorithm at stationarity in direction $j$, while the denominator is the squared width of the bulk of the posterior in direction $j$. Consequently, $\gamma$ will be close to 1 whenever the average step size at stationarity is small relative to the width of the bulk of the posterior. 

The purpose of CDA is to introduce additional parameters that allow us to control the step size relative to the posterior width -- roughly speaking, the ratio in \eqref{eq:missinginfo} -- with greater flexibility than reparametrization or parameter expansion. The flexibility gains are achieved by allowing the invariant measure to change as a result of the introduced parameters. The additional parameters, which we denote $(r,b)$, correspond to a collection of reparametrizations, each of which defines a proper (but distinct) likelihood $L_{r,b}(\theta;y)$, and for which there exists a Gibbs update rule of the form \eqref{eq:da}. In general, $b$ will correspond to a location parameter and $r$ a scale parameter that are tuned to increase $\bb E[\var(g(\theta) \mid z)]\{\var(g(\theta))\}^{-1}$, although the exact way in which they enter the likelihood and corresponding Gibbs update depend on the application. The reparametrization also has the property that $L_{1,0}(\theta;y) = L(\theta;y)$, the original likelihood. The resulting Gibbs sampler, which we refer to as CDA Gibbs, has $\theta$-marginal invariant measure $\Pi_{r,b}(\theta;y) \propto L_{r,b}(\theta;y) \Pi^0(\theta)$, where $\Pi^0(\theta)$ is the prior. Ultimately, we are interested in $\Pi_{1,0}(\theta;y)$, so we use CDA Gibbs as an efficient proposal for Metropolis-Hastings. That is, we propose $\theta^*$ from $Q(\theta;\cdot)$ where
\be \label{eq:Q}
Q_{r,b}(\theta;A) = \int_{(\theta^*,z) \in A \times \mc Z} \pi_{r,b}(z;\theta,y) f_{r,b}(\theta^*;z,y) dz d\theta^*
\ee
for $A \subseteq \Theta$, where $\pi_{r,b}$ and $f_{r,b}$ denote the conditional densities of $z$ and $\theta$ in the Gibbs sampler with invariant measure $\Pi_{r,b}$. By tuning working parameters during an adaptation phase to reduce the autocorrelations and increase the Metropolis-Hastings acceptance rate, we can select values of the working parameters that yield a computationally efficient algorithm. Tuning is facilitated by the fact that the M-H acceptance ratios using this proposal kernel have a convenient form, which is a nice feature of using Gibbs to generate M-H proposals.
\begin{remark} \label{rem:accrat}
The CDA M-H acceptance ratio is given by
\be
1 \wedge \frac{L(\theta';y) \Pi^0(\theta') Q_{r,b}(\theta;\theta')}{L(\theta;y) \Pi^0(\theta) Q_{r,b}(\theta';\theta)} = 1 \wedge \frac{L(\theta';y)L_{r,b}(\theta;y)}{L(\theta;y)L_{r,b}(\theta';y)} \label{eq:mh-accrat}
\ee
\end{remark}
A general strategy for tuning is given in Section \ref{sec:tuning}. 

We give a basic convergence guarantee that holds for the CDA M-H under weak assumptions on $L_{r,b}$, which is based on \cite[Theorem 3, also pp. 214]{roberts1994simple}. Basically, one needs $\Pi(\cdot) \ll \Pi_{r,b}(\cdot)$ for all $r,b$, where for two probability measures $\mu,\nu$, $\mu(\cdot) \ll \nu(\cdot)$ means $\mu$ is absolutely continuous with respect to $\nu$.
\begin{remark}[Ergodicity] \label{rem:ergodic}
Assume that $\Pi(d\theta)$ and $\Pi_{r,b}(d\theta)$ have densities with
respect to Lebesgue measure on $\bb R^p$, and that \\ $K_{r,b}((\theta,z);(\theta',z'))>0 \,\forall\, ((\theta,z),(\theta',z')) \in (\Theta \times \mc Z) \times (\Theta \times \mc Z)$. Then, 
\begin{itemize}
\item For fixed $r,b$, CDA Gibbs is ergodic with invariant measure $\Pi_{r,b}(d\theta,dz)$.
\item A Metropolis-Hastings algorithm with proposal kernel $Q_{r,b}(\theta';\theta)$ as defined in \eqref{eq:Q} with fixed $r,b$ is ergodic with invariant measure $\Pi(d\theta)$.
\end{itemize}
\end{remark}
Proofs are located in the Appendix. 
%Remark \ref{rem:ergodic} provides a convergence guarantee, but does not motivate why the convergence properties of CDA are superior to the original Gibbs. Because the algorithms we propose differ in the details of how $r,b$ enter $K$, a general result on convergence rates is not possible. However, we do give several heuristics for why CDA is expected to result in faster convergence in Section \ref{sec:algos}. 

\subsection{Initial Example: Probit with Intercept Only}
We use a simple example to illustrate CDA. Consider an intercept-only probit
\be
y_i \sim \Bern(p_i), \quad p_i = \Phi(\theta) \quad i=1,\ldots,n
\ee
and improper prior $\Pi^0(\theta) \propto 1$. The basic data augmentation algorithm \citep{tanner1987calculation,albert1993bayesian} has the update rule
\be
z_i \mid \theta, y_i &\sim \left\{ \begin{array}{cc} \No_{[0,\infty)}( \theta,1) & \text{ if } y_i = 1 \\ \No_{(-\infty,0]}( \theta,1) & \text{ if } y_i = 0 \end{array} \right. \quad i=1,\ldots,n\\
\theta \mid z, y &\sim \No\left( n^{-1} \sum_i z_i, n^{-1} \right),
\ee
where $\No_{[a,b]}(\mu,\sigma^2)$ is the normal distribution with mean $\mu$ and variance $\sigma^2$ truncated to the interval $[a,b]$. \cite{johndrow2016inefficiency} show that when $\sum_i y_i = 1$, $\var(\theta_t \mid \theta_{t-1})$ is approximately $n^{-1} \log n$, while the width of the high probability region of the posterior is order $(\log n)^{-1}$, leading to slow mixing. 

As the conditional variance $\mbox{var}(\theta\mid z,y)$ is independent of $z$, we introduce a scale parameter $r$ in the update for $z$, then adjust the conditional mean by a location parameter $b$.  This is equivalent to changing the scale of $z_i \mid\theta,y_i$ from $1$ to $r$ and the mean from $\theta$ to $\theta+b$. These adjustments yield 
\be
\mbox{pr}(y_i = 1 | \theta, r, b) =& \int_{0}^{\infty} \frac{1}{\sqrt{2 \pi r} } \exp\left(-\frac{(z_i-\theta-b)^2}{2 r^2} \right) dz_i 
\\ = & \Phi\bigg( \frac{\theta+b}{\sqrt{r}}\bigg),
\label{eq:prop-marginal-probit-intercept}
\ee
leading to the modified data augmentation algorithm
\be \label{eq:cda-probit-intercept}
z_i \mid \theta, y_i &\sim \left\{ \begin{array}{cc} \No_{[0,\infty)}( \theta+b,r) & \text{ if } y_i = 1 \\ \No_{(-\infty,0]}( \theta+b,r) & \text{ if } y_i = 0 \end{array} \right.  \quad i=1,\ldots,n\\
\theta \mid z,y &\sim \No\left(n^{-1}  \sum_i(z_i-b), n^{-1} r \right).
\ee

To achieve step sizes consistent with the width of the high posterior probability region, we need $n^{-1} r \approx (\log n)^{-1}$, so $r \approx n/\log n$. To preserve the original target, we use (\ref{eq:cda-probit-intercept}) to generate an M-H proposal $\theta^*$. By Remark \ref{rem:accrat}, the M-H acceptance probability is given by \eqref{eq:mh-accrat} with $L_{r,b}(\theta;y_i) =  \Phi\big( ({\theta+b}){r}^{-1/2}\big) ^{y_i} \Phi\big( - ({\theta+b}){r}^{-1/2}\big)^{(1-y_i)}$ and $L(\theta;y_i)  = L_{1,0}(\theta;y_i)$. Setting $r_i=1$ and $b_i=0$ leads to acceptance rate of $1$, which corresponds to the original Gibbs sampler.

To illustrate, we consider  $\sum_i y_i =1$ and $n=10^4$. Letting $r = n/\log n$, we then choose the $b_i$'s to increase the acceptance rate in the M-H step. In this simple example, it is easy to compute a ``good'' value of $b_i$, since $b_i = -3.7 (\sqrt r -1)$ results in $\mbox{pr}(y_i = 1) = \Phi(-3.7) = n^{-1}\sum_i y_i  \approx 10^{-4}$ in the proposal distribution, centering the proposals near the MLE for $p_i$.

\begin{figure}[H]
  {\caption{Autocorrelation functions (ACFs) and kernel-smoothed density estimates for different CDA samplers in intercept-only probit model.}}
  {%
    \subfigure[ACF for CDA without M-H adjustment.]{\label{probit_demo_intercept_proposal}
      \includegraphics[width=0.3\linewidth]{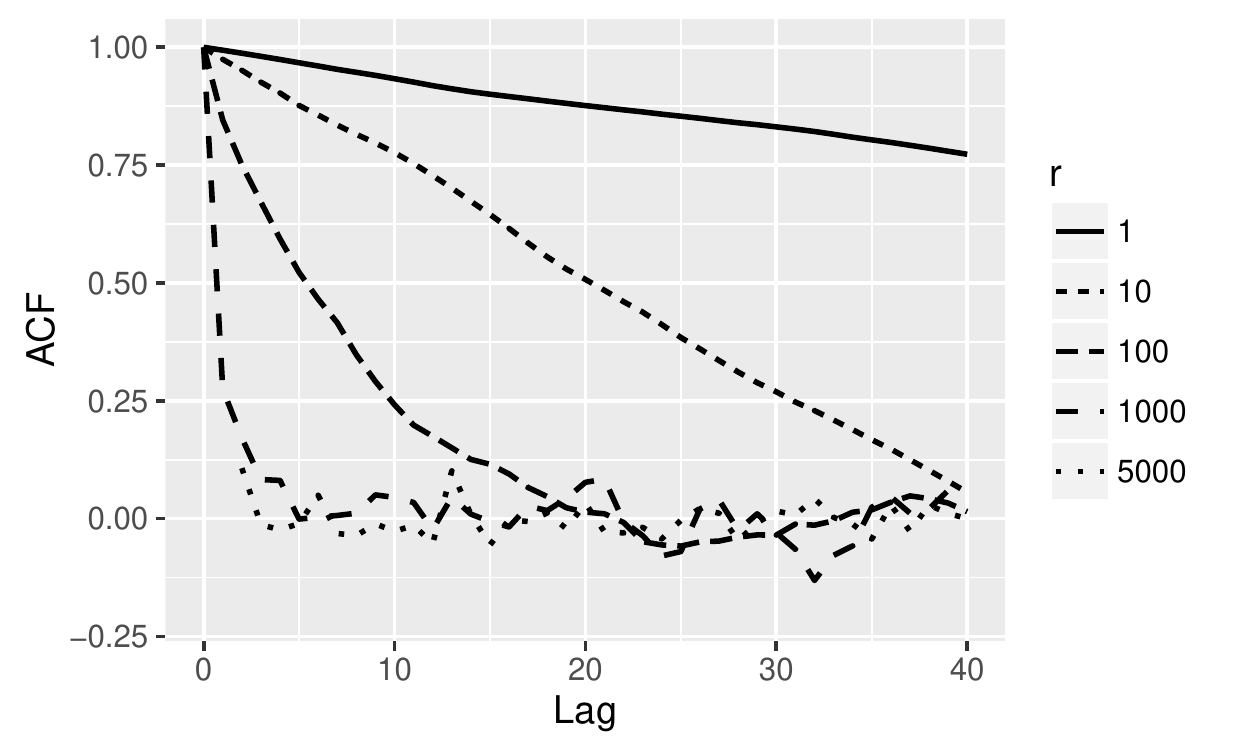}}%
    \quad
    \subfigure[Posterior density estimates without M-H adjustment.]{\label{probit_demo_intercept_density}
      \includegraphics[width=0.3\linewidth]{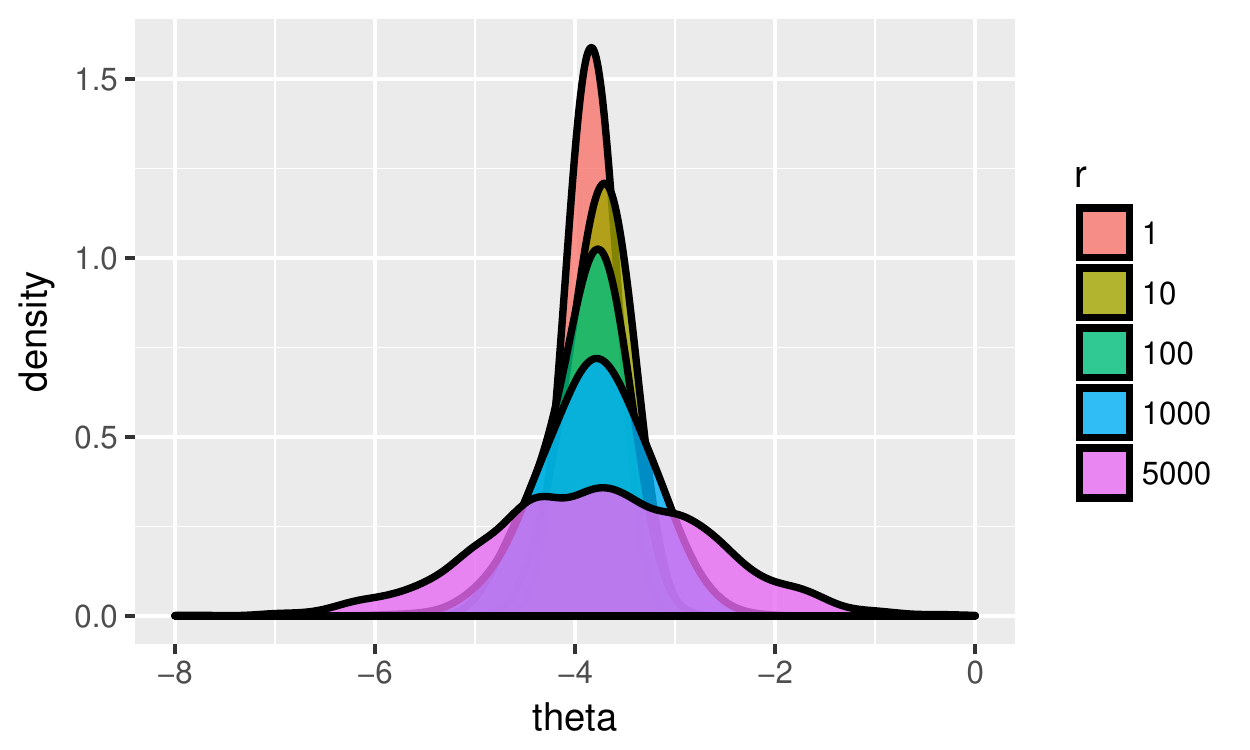}}
    \quad
    \subfigure[ACF for CDA with M-H adjustment]{\label{probit_demo_intercept_posteriorsample}
      \includegraphics[width=0.3\linewidth]{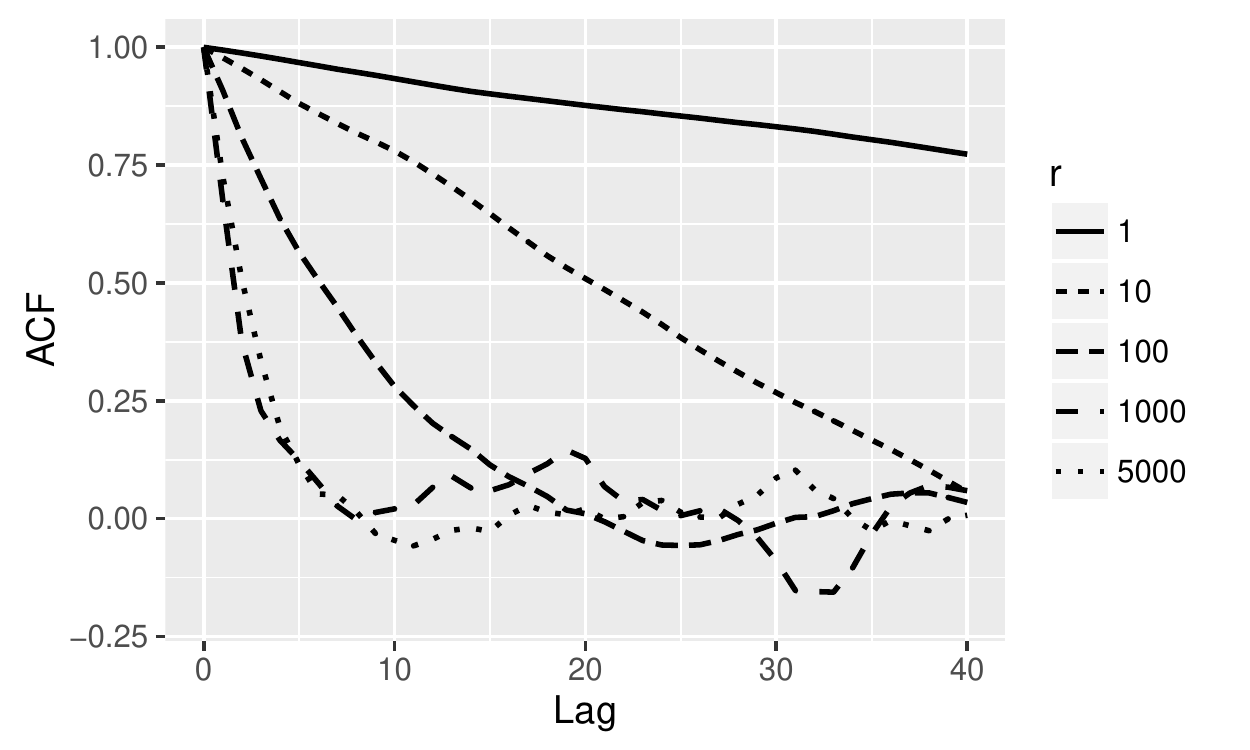}}
  }
   \label{probit_demo_intercept}
\end{figure}

We perform computation for these data with different values of $r$ ranging from $r=1$ to $r=5,000$, with $r=1,000 \approx n/\log n$ corresponding to the theoretically optimal value.  Figure~\ref{probit_demo_intercept_proposal} plots autocorrelation functions (ACFs) for these different samplers without  M-H adjustment. Autocorrelation is very high even at lag 40 for $r=1$, while increasing $r$ leads to dramatic improvements in mixing. There are no further gains in increasing $r$ from the theoretically optimal value of $r=1,000$ to $r=5,000$. Figure~\ref{probit_demo_intercept_density} shows kernel-smoothed density estimates of the posterior of $\theta$ without M-H adjustment for different values of $r$ and based on long chains to minimize the impact of Monte Carlo error; the posteriors are all centered on the same values but with variance increasing somewhat with $r$.  With M-H adjustment such differences are removed; the M-H step has acceptance probability close to one for $r=10$ and $r=100$, about 0.6 for $r=1,000$, and 0.2 for $r=5,000$.

\section{Specific Algorithms} \label{sec:algos}
In this section, we describe CDA algorithms for general probit and logistic regression, and describe a general strategy for tuning $r,b$.

\subsection{Probit Regression}
Consider the probit regression:
\be
y_i \sim \Bern(p_i), \quad p_i = \Phi(x_i \theta)  \quad i=1,\ldots,n
\ee
with improper prior $\Pi^0(\theta) \propto 1$. The data augmentation sampler \citep{tanner1987calculation, albert1993bayesian} has the update rule
\be
z_i \mid \theta, x_i, y_i &\sim \left\{ \begin{array}{cc} \No_{[0,\infty)}(x_i \theta,1) & \text{ if } y_i = 1 \\ \No_{(-\infty,0]}(x_i \theta,1) & \text{ if } y_i = 0 \end{array} \right.  \quad i=1,\ldots,n\\
\theta \mid z, x, y &\sim \No((X'X)^{-1} X'z, (X'X)^{-1}).
\ee
\cite{liu1999parameter} and \cite{meng1999seeking}, among others, previously studied this algorithm and proposed to rescale $\theta$ through parameter expansion. However, this modification does not impact the conditional variance of $\theta$ and thus does not directly increase typical step sizes.

Our approach is fundamentally different, since we directly adjust the conditional variance. Similar to the intercept only model, we modify $\mbox{var} (\theta| z)$ by changing the scale of each $z_i$. Since the conditional variance is now a matrix, for flexible tuning, we let $r$ and $b$ vary over index $i$, yielding update rule
\be \label{eq:cda-probit}
z_i \mid \theta, x_i, y_i &\sim \left\{ \begin{array}{cc} \No_{[0,\infty)}(x_i \theta+b_i,r_i) & \text{ if } y_i = 1 \\ \No_{(-\infty,0]}(x_i \theta+b_i,r_i) & \text{ if } y_i = 0 \end{array} \right.  \quad i=1,\ldots,n\\
\theta \mid z, X &\sim \No((X'R^{-1}X)^{-1} X'R^{-1}(z-b), (X'R^{-1}X)^{-1}),
\ee
where $R = \diag(r_1,\ldots,r_n)$, $b = (b_1,\ldots,b_n)'$, under the Bernoulli likelihood:
\be
\mbox{pr}(y_i = 1 | \theta, x_i, r_i, b_i) = & \int_{0}^{\infty}
\frac{1}{\sqrt{2 \pi r_i} } \exp\left(-\frac{(z_i-x_i\theta-b_i)^2}{2 r_i}
\right) dz_i \\
= & \Phi\bigg( \frac{x_i\theta+b_i}{\sqrt{r_i}}\bigg).
\label{eq:prop-marginal-probit}
\ee
For fixed $r = (r_1,\ldots,r_n)$ and $b = (b_1,\ldots,b_n)$, \eqref{eq:prop-marginal-probit} defines a proper Bernoulli likelihood for $y_i$ conditional on parameters, and therefore the transition kernel $K_{r,b}((\theta,z);\cdot)$ defined by the Gibbs update rule in \eqref{eq:cda-probit} would have a unique invariant measure for fixed $r,b$, which we denote $\Pi_{r,b}(\theta,z \mid y)$. 
%\eqref{eq:prop-marginal-probit} suggests that another way to view CDA Gibbs is that a different link function is chosen for each observation, and the corresponding M-H algorithm allows us to recover the posterior under the original link. The CDA M-H algorithm in this case proposes from $Q(\theta^*;\theta) = \int f(\theta^*;z,y)  \pi(z;\theta,y) dz$, the $\theta$-marginal of the transition kernel $K_{r,b}((\theta^*,z^*);(\theta,z))$. The acceptance probability is given by \eqref{eq:mh-accrat} with $L_{r,b}(\eta_i;y_i) = \Phi\big( \frac{\eta_i+b}{\sqrt{r}}\big) ^{y_i} \Phi \big( -\frac{\eta_i+b}{\sqrt{r}}\big) ^{(1-y_i)}$; we denote $L_{1,0}$ by $L$. 

For insight into the relationship between $r$ and step size, consider the $\theta$-marginal autocovariance in a Gibbs sampler evolving according to $K_{r,b}$:
\be
\cov_{r,b}(\theta_t \mid \theta_{t-1},X,z,y) &= (X'R^{-1}X)^{-1} \\ & + (X'R^{-1}X)^{-1} X'R^{-1}\cov(z-b | R) R^{-1}X(X'R^{-1}X)^{-1} \\
&\ge (X'R^{-1}X)^{-1}, \label{eq:varlb-probit}
\ee
In the special case where $r_i = r_0$ for all $i$, we have
\be
\cov_{r,b}(\theta_t \mid \theta_{t-1}, X,z,y) &\ge r_0 (X'X)^{-1}, 
\ee
so that all of the conditional variances are increased by at least a factor of $r_0$. This holds uniformly over the entire state space, so it follows that 
\be
\bb E_{\Pi_{r,b}}[\var(\theta_j \mid z)] \ge r_0 \bb E_{\Pi}[\var(\theta_j \mid z)]. 
\ee
The key to CDA is to choose $r,b$ to make $\bb E_{\Pi_{r,b}}[\var(\theta_j \mid z)]$ close to $\var_{\Pi_{r,b}}(\theta_j \mid z)$, while additionally maximizing the M-H acceptance probability. We defer the choice for $r,b$ and their effects to the last subsection.

\subsection{Logistic Regression}
Calibration was easy to achieve in the probit examples, because $\mbox{var}( \theta | z,y)$ does not involve the latent variable $z$.  In cases in which the latent variable impacts the variance of the conditional posterior distribution of $\theta$, we propose to stochastically increase $\mbox{var}(\theta|z,y)$ by modifying the distribution of $z$. We focus on the logistic regression model with 
\be
y_i \sim \Bern(p_i), \quad p_i = \frac{\exp(x_i \theta)}{1+\exp(x_i \theta)} \quad i=1,\ldots,n
\ee
and improper prior $\Pi^0(\theta) \propto 1$. For this model, \cite{polson2013bayesian} proposed Polya-Gamma data augmentation:
\be
 z_i &\sim {\PG}(1, |\xtheta|) \quad i=1,\ldots,n,\\
\theta &\sim \No \left(  (X' Z X)^{-1}   X'  (y-0.5)  ,  (X' Z X)^{-1}  \right),
\ee
where $Z= \diag(z_1,\ldots,z_n)$.  This algorithm relies on expressing the logistic regression likelihood as
$$L( \xtheta; y_i )=  \int \exp\{ \xtheta (y_i-1/2)\} \exp\bigg\{ -\frac{z_i (\xtheta)^2}{2}\bigg\} \PG(z_i \mid 1,0) dz_i,$$
where $\mbox{PG}(a_1,a_2)$ denotes the Polya-Gamma distribution with parameters $a_1,a_2$, with $\bb{E}z_i= {a_1}/{(2 a_2)}\tanh({a_2}/{2})$.

We replace $\PG(z_i \mid 1,0)$ with $\PG(z_i \mid r_i,0)$ in the step for updating the latent data. Since $\bb E_z \mbox{var}(\theta|z,y)$ lacks closed-form, we focus on the precision matrix $\bb E_z \big( \mbox{var}(\theta|z,y)\big)^{-1} = X' \bb E Z X$. Smaller $r_i$ can lead to smaller $\bb E z_i$, providing a route to calibration. Applying the bias-adjustment term $b_i$ to the linear predictor $\eta_i = x_i\theta$ leads to 
\be
L_{r,b}(\xtheta;y_i) = & \int_{0}^{\infty}  \exp\{ (\xtheta+b_i)
(y_i-r_i/2)\} \\ & \quad \exp\bigg\{ -\frac{z_i (\xtheta+b_i)^2}{2}\bigg\} \PG(z_i \mid r_i,0) dz_i \\
= &  \frac{\exp \{ (x_i \theta + b_i)y_i \}}{\{1+\exp(\xtheta +b_i)\}^{r_i}},
\label{eq:prop-marginal-logit}
\ee
and the update rule for the CDA Gibbs sampler is then 
\be
 z_i &\sim {\PG}(r_i, |\xtheta+b_i|) \quad i=1,\ldots,n,\\
\theta^* &\sim \No \left(  (X' Z X)^{-1}  X'  (y -r/2- Zb) ,  (X' Z X)^{-1}  \right),
\ee
where $r = (r_1,\ldots,r_n)$. By \eqref{eq:mh-accrat}, the M-H acceptance probability is
\be
1 \wedge \prod_i  \frac{ \{1+\exp(\xtheta)\}   \{1+\exp(\xtheta^*+b_i)\}^{r_i} } {  \{1+\exp(\xtheta^*)\}  \{1+\exp(\xtheta+b_i)\}^{r_i}    }.
\ee

\subsection{Choice of Calibration Parameters} \label{sec:tuning}
As illustrated in the previous subsection, efficiency of CDA is dependent on a good choice of the calibration parameters $r=(r_1,\ldots,r_n)$ and $b=(b_1,\ldots,b_n)$.  We propose a simple and efficient algorithm for calculating ``good'' values of these parameters relying on Fisher information.  {Although our choice of calibration parameters relies on large data sample arguments, we find that this calibration approach also works well in smaller data samples.}

{
Our
goal is to adjust the conditional variance under calibration of $(r,b)$ to approximately match the marginal variance under the exact target distribution.
{
The inverses of the following Fisher information provide  useful approximation to the two posterior covariances.

\be   \left (\mc I_{y\mid \theta}({\theta}) \right)_{i,j} &  = \bb E_{y\mid \theta} \left[ \left( \frac{\partial}{\partial \theta_i} \log L(y;\theta) \right) \left( \frac{\partial}{\partial \theta_j} \log L(y;\theta) \right) \right], \\
 \left(  \mc I_{y\mid \theta,z}({\theta};r,b) \right)_{i,j}& = \bb E_{y\mid \theta,z} \left[ \left( \frac{\partial}{\partial \theta_i} \log L_{r,b}(y,z;\theta) \right)\left( \frac{\partial}{\partial \theta_j} \log L_{r,b}(y,z;\theta) \right) \right]
\ee
}
for $i=1,\ldots,p$, $j=1,\ldots,p$, with $\bb E_{y\mid \theta}$ taken over the distribution of $y$ under the target marginal $L(y;\theta)$ and $\bb E_{y\mid \theta,z}$ taken over the conditional distribution of $y$ under the augmented $L_{r,b}(y,z;\theta)$ with the calibration of $(r,b)$.
Since $\mc I_{y\mid \theta,z}({\theta};r,b)$ depends on random $z$, we marginalize over
the conditional distribution
of $z$ under $L_{r,b}(\theta; y)$ and obtain $\bb E_{z\mid \theta}\mc I_{y\mid \theta,z}(\theta;r,b)$. Via adjusting $r$, one can then  minimize the difference between $\mc  I_{y\mid \theta}(\theta)$
and $\bb E_{z\mid \theta}\mc I_{y\mid \theta,z}(\theta;r,b)$.}

{
Often, one can avoid computing the full Fisher information. For  each class of models under the same data augmentation, they share the same form of conditional likelihoods for $y\mid \eta(\theta)$ given a mapping $\eta(\theta):\bb
R^p\rightarrow \bb R^d $.
 For example,  all Bernoulli probit models  follow $y_{i}\mid \eta_{i}(\theta) \stackrel{iid}{\sim}\Bern( \Phi(\eta_{i}(\theta)) )$, except with different  $\eta(\theta)$. The above full Fisher information can be rewritten as
 \be\mc I_{y\mid \theta}({\theta})  &  =  \dot \eta\mc I_{y\mid \theta}(\eta({\theta}))  \dot \eta', \\
  \mc I_{y\mid \theta,z}({\theta};r,b) & = \dot \eta\mc I_{y\mid \theta,z}(\eta({\theta)};r,b)  \dot \eta',
\ee
where $\dot \eta$ denotes the $p$-by-$d$ gradient matrix consisting of the partial
derivative ${\partial
\eta_{k}(\theta)}/{\partial \theta_j}$ of  the $k$th output 
$\eta_{k}(\theta)$ with respect to $\theta_j$. It suffices to reduce the difference between  $\mc I_{y\mid \theta}(\eta({\theta}))$  and $\mc I_{y\mid \theta,z}(\eta({\theta)};r,b)$ instead of the full Fisher
 information. The solution is a function of $\eta$, with form invariant to models under the same conditional likelihood of $y\mid \eta$.
 }

{
In all CDA algorithms presented in
this article,  $\mc I_{y\mid \theta}(\eta({\theta}))$  and  $\mc I_{y\mid \theta,z}(\eta({\theta)};r,b)$ are simple diagonal matrices, and 
 it can be made exactly $\mc I_{y\mid \theta}({\theta})=\bb E_{z\mid \theta}\mc I_{y\mid \theta,z}({\theta};r,b)$ for given $\theta$ with a closed-form
solution. As there could be more complicated
scenarios, we suggest the following. When the difference cannot
be simply eliminated, one could utilize a metric between two matrices, such
as Rao's distance $\{\text{tr}
[\log(A^{-1/2}BA^{-1/2})^{2}]\}^{1/2}$ with $\text{tr}$ as the trace \citep{atkinson1981rao},
and an optimization algorithm to minimize the difference. When the Fisher information
is intractable to compute, one could instead utilize the observed Fisher information
\citep{efron1978assessing},    
$$\bigg(\hat{\mc I}_{y\mid \theta}({\theta}) \bigg)_{i,j}   =  \bigg( \frac{\partial}{\partial \theta_i} \log L(y;\theta) \bigg) \bigg( \frac{\partial}{\partial \theta_j} \log L(y;\theta) \bigg),$$ 
$$ \bigg(  \hat{\mc I}_{y\mid \theta,z}({\theta};r,b) \bigg)_{i,j} = \bigg( \frac{\partial}{\partial \theta_i} \log L_{r,b}(y,z;\theta) \bigg)\bigg( \frac{\partial}{\partial \theta_j} \log L_{r,b}(y,z;\theta) \bigg),$$
with $y$ the observed data.
}

{As the Fisher information matrices depend on $\theta$, we use  an adaptation phase to dynamically update $r_t,b_t$ with posterior sample $\theta_t$. Then
we stop adaptation after $\theta_t$ enters the high posterior
density region. This approach is similar to using the frequentist Fisher information evaluated at the maximum-a-posteriori
(MAP) estimate, except it does not require the extra optimization steps  for computing
the MAP. It works well empirically in examples we have considered.}

 Specifically, we choose $r_{t+1}$ to minimize the  difference between $\mc  I_{y\mid \theta_{t}}(\theta_{t})$
and  $\bb E_{z\mid \theta_{t}}\mc I_{y\mid \theta_{t},z}(\theta_{t};r_{t+1},b_{t})$, or $\mc  I^{-1}_{y\mid \theta_{t}}(\theta_{t})$
and $\bb E_{z\mid \theta_{t}}\mc I^{-1}_{y\mid \theta_{t},z}(\theta_{t};r_{t+1},b_{t})$.
Additionally, we set $b_{t+1}$ to minimize the difference between
$L_{1,0}(\theta_t;y)$ and  $L_{r_{t+1},b_{t+1}}(\theta_t;y)$. Thus, we use $r$ to adjust the conditional variance based on $L_{r,b}$ to match the marginal variance based on $L$ and $b$ to make $L_{r,b}$ close to $L_{1,0}$ in the neighborhood of $\theta_t$. Intuitively, this will make the target distribution closer to the invariant measure of calibrated Gibbs, and correspondingly increase the MH acceptance rate. {Some illustrative results about the adaptation
are provided in the appendix.} The proposal kernel we describe above is \emph{adaptive}; that is, we have a collection of proposal kernels $\mc Q = \{Q_{r,b}\}_{(r,b) \in \bb R_+ \times \bb R}$, and we choose a different member of $\mc Q$ at each iteration to create the proposal. 
%The \emph{target} for the resulting transition kernel is $\Pi_{1,0}$ for every $Q_{r,b}$ because of the Metropolis-Hastings rejection step.  
In general, ergodicity of adaptive algorithms requires a diminishing adaptation condition \citep{roberts2007coupling}.  {For simplicity, we satisfy this condition by stopping adaptation after a tuning phase.}

For a concrete illustration, we first return to the first example of probit regression. Letting $\eta_i = x_i\theta$, we obtain
\be
\mc I_{y\mid \theta}({\theta}) &=  \dot\eta\diag\bigg\{\frac{\phi(\eta_i)^2}{ {\Phi(\eta_i)(1- \Phi(\eta_i))}}\bigg\}\dot\eta',\qquad \bb E_{z\mid \theta}\mc I_{y\mid \theta,z}({\theta};r,b)= \dot\eta R^{-1} \dot\eta',
\ee
 where $\phi$ is the standard normal density, with $\dot\eta=X'$. {Having $\mc I_{y\mid \theta}({\theta})= \bb
 E_{z\mid \theta}\mc I_{y\mid \theta,z}({\theta};r,b)$  and $L_{r,b}(\eta_i;y_i)= L(\eta_i;y_i)
$  yields 
\be
r_i &= \frac{\Phi(\eta_i)(1- \Phi(\eta_i))} {\phi(\eta_i)^2},\\
b_i &= \eta_i (\sqrt{r_i}-1).
\ee
 For Bernoulli probit models with other forms of $\eta_i$, the solution for tuning parameters remains the same.}

\begin{figure}[H]
  {%
    \subfigure[Traceplot for the original DA, parameter expanded DA and CDA algorithms.]{\label{probit_reg_trace}%
      \includegraphics[width=0.45\linewidth]{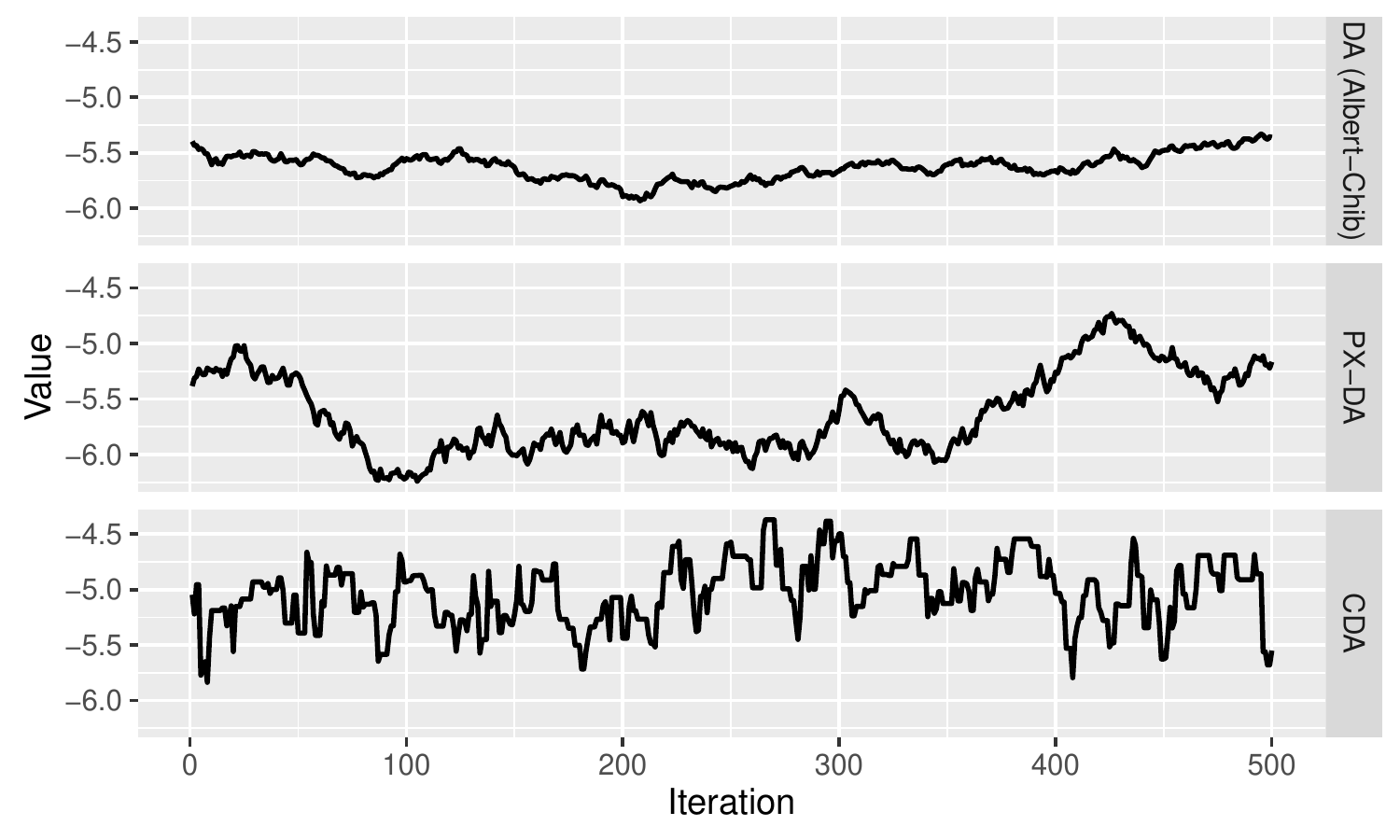}}%
    \qquad
    \subfigure[ACF for original DA, parameter expanded DA and CDA algorithms.]{\label{probit_reg_acf}%
      \includegraphics[width=0.45\linewidth]{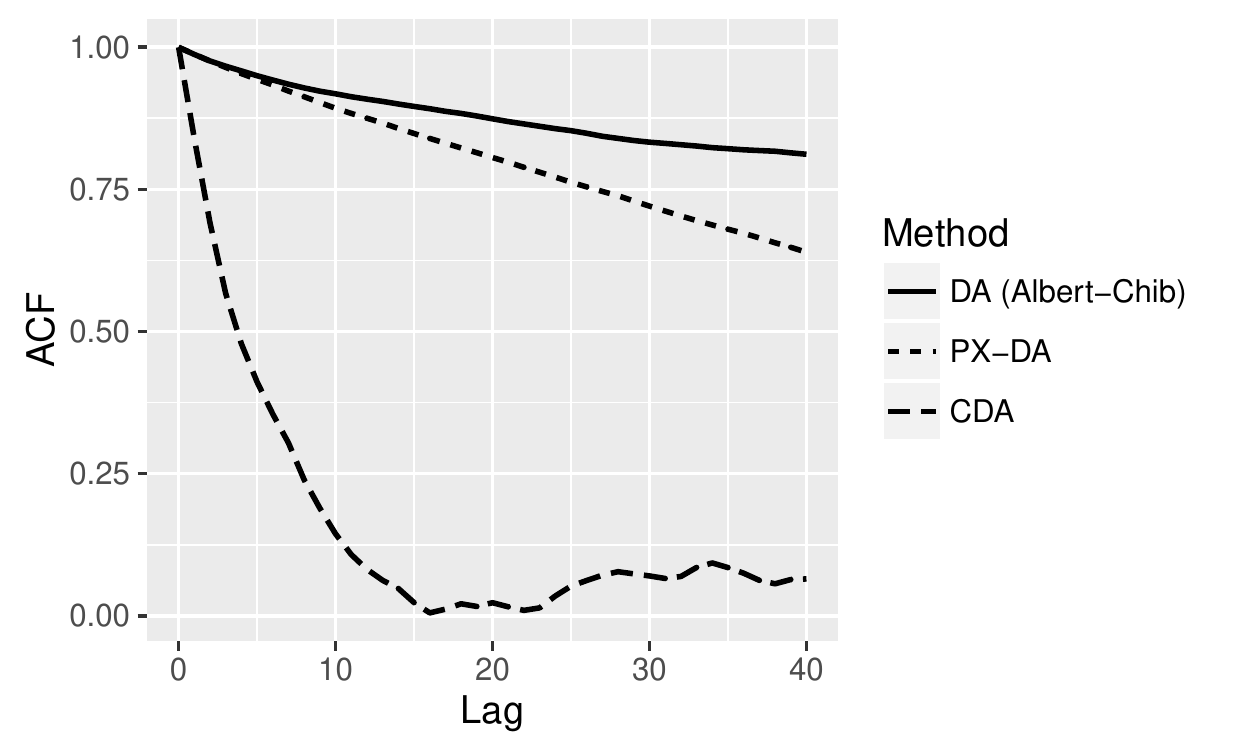}}
  }
 {\caption{Panel (a) demonstrates in traceplot and panel (b) in autocorrelation the substantial improvement in CDA by correcting the variance mis-match in probit regression with rare event data, compared with the original \citep{albert1993bayesian} and parameter-expanded methods \citep{liu1999parameter}.}}
\end{figure}

 In simulation, we consider a probit regression with an intercept and two predictors $x_{i,1},x_{i,2}\sim \No(1,1)$, with $\theta=(-5,1,-1)'$, generating $\sum y_i=20$ among $n=10,000$. The \cite{albert1993bayesian} DA algorithm mixes slowly (Figure~\ref{probit_reg_trace} and \ref{probit_reg_acf}). We also show the 
results of the parameter expansion algorithm (PX-DA) proposed by \cite{liu1999parameter}. PX-DA only mildly reduces the correlation, as it does not solve the small step size problem.  For CDA, we tuned $r$ and $b$ for $100$ steps using the Fisher information, reaching a satisfactory acceptance rate of $0.6$ {and leading to dramatically better mixing}.

For the second example of logistic regression, taking $\eta_i=x_i\theta$, the Fisher information matrices
are:
 \begin{eqnarray}
\mc I_{y\mid \theta}({\theta}) =&  \dot\eta \diag\bigg\{\frac{\exp(\eta_i)}{ \{1+\exp(\eta_i)\} ^2}\bigg\} \dot\eta', \nonumber \\
\bb E_{z\mid \theta}\mc I_{y\mid \theta,z}({\theta};r,b)= & \dot\eta  \diag\bigg\{ \frac{r_i}{2 |\eta_i+b_i|}\tanh\Big(\frac{|\eta_i+b_i|}{2} \Big)\bigg\} \dot\eta' .  \nonumber
\end{eqnarray}
where $\dot\eta=X'$. 
Setting $\mc I_{y\mid \theta}({\theta})=\bb E_{z\mid \theta}\mc I_{y\mid \theta,z}({\theta};r,b)$ and $ \{1+\exp(\eta_i)\}  = \{1+\exp(\eta_i+b_i)\}^{r_i}$
to locally maximize the M-H acceptance rate yields
\be
r_i & =\frac{\exp(\eta_i)}{ \{1+\exp(\eta_i)\} ^2} {2 |\eta_i+b_i|}/ \tanh\Big(\frac{|\eta_i+b_i|}{2} \Big),
\\ b_i & = \log[  \{1+\exp(\eta_i)\}^{1/r_i} -1] - \eta_i.
\ee
{Again, this tuning solution is invariant to the different forms of $\eta$ under
the Bernoulli logistic model.}

\begin{figure}[H]
  {%
    \subfigure[Traceplots for DA, CDA and M-H with multivariate normal proposal.]{%
      \includegraphics[width=0.45\linewidth]{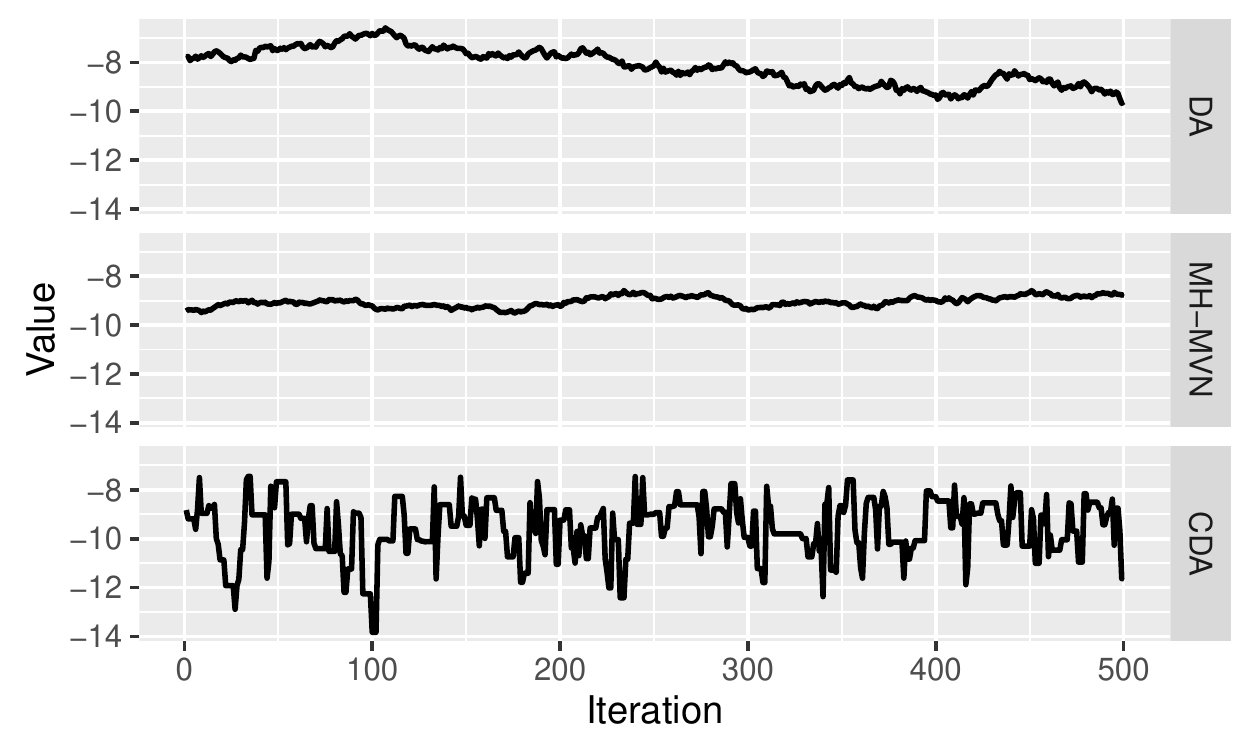}}%
    \qquad
    \subfigure[ACF for DA, CDA and M-H with multivariate normal proposal.]{%
      \includegraphics[width=0.45\linewidth]{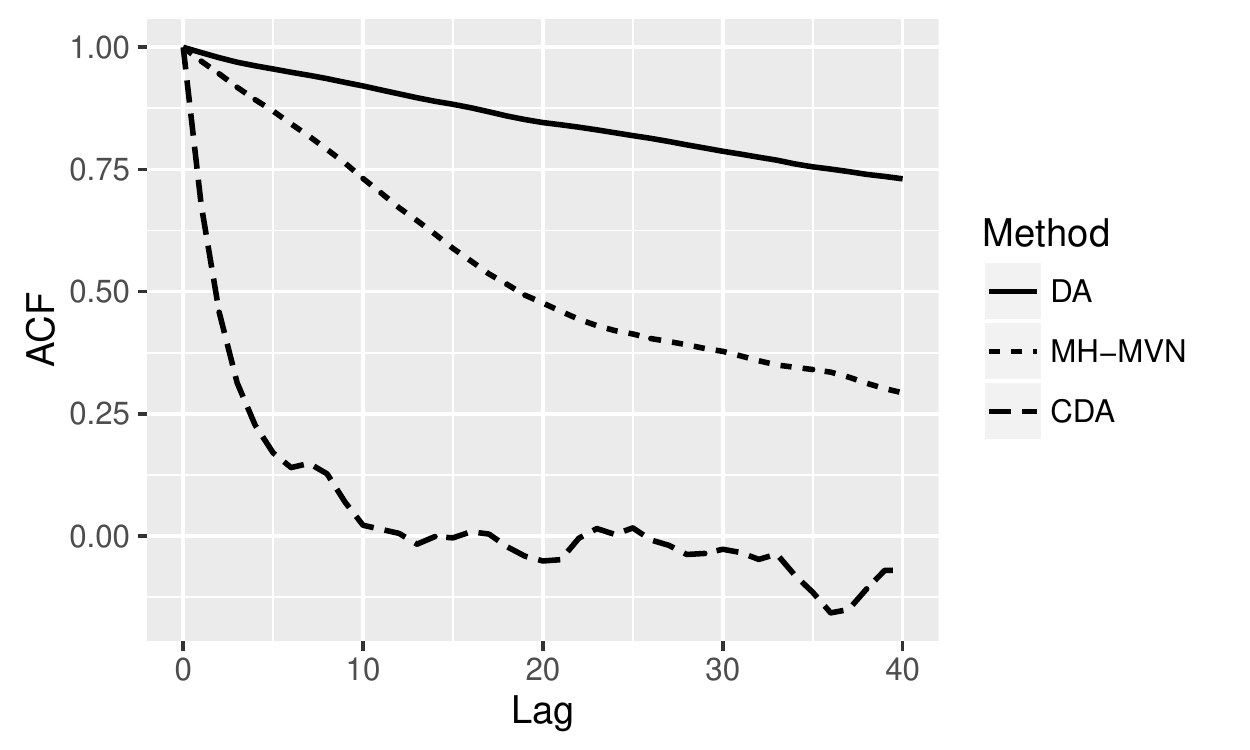}}
  }
  {\caption{Panel (a) demonstrates in traceplot and panel (b) in autocorrelation the substantial improvement of CDA in logistic regression with rare event data, compared with the original DA \citep{polson2013bayesian} and the M-H algorithm with multivariate normal proposal (MH-MVN).\label{logit_random_mixing}}}
\end{figure}

To illustrate, we use a two parameter intercept-slope model with $x_1\sim \No(0,1)$ and $\theta=(-9,1)'$. With $n= 10^5$, we obtain rare outcome data with 
$\sum y_{i} = 50 $.  Besides the original DA algorithm  \citep{polson2013bayesian}, we also consider an M-H sampler using a multivariate normal proposal $\theta^*|\theta \sim \No(\theta^*| \theta, {\mc I}^{-1}(\theta))$ with the inverse Fisher information as the covariance. {Similarly, we test an alternative of using DA to generate new $\theta^*$, and scaling to $\theta^{**}=\theta+\alpha(\theta^{*}-\theta)$, with $\alpha\ge 1$, as an M-H proposal. Both M-H with a normal proposal and with scaled proposal suffer from low acceptance rate, unless $\alpha\approx 1$ in the latter (corresponding to almost no adjustment from DA).} For CDA we tuned $r$ and $b$ for $100$ steps, reaching an acceptance rate of $0.8$.  Shown in Figure~\ref{logit_random_mixing}, DA and simple M-H mix slowly, exhibiting strong autocorrelation even at lag $40$, while CDA has dramatically better mixing.

{
\section{Simulation Study: Scaling to Massive $n$}
}

As motivated above,  two factors are necessary for obtaining usable posterior samples within a practical  time: a low computing cost  in each iteration and a high effective sample size within a small number of iterations. We first demonstrate that calibration can solve the latter issue.

To start, we consider a simple Bernoulli logistic regression with a common intercept:

$$y_i\stackrel{iid}{\sim} \Bern\Big(\frac{\exp(\theta)}{1+\exp(\theta)}\Big), \quad i=1,\ldots,n,$$
with a flat improper prior for $\theta$. As the likelihood is \\ $L(y;\theta )= {\exp(\theta\sum y_i)}{(1+\exp(\theta))^{-n}} $, it enjoys efficient computing per iteration
that only involves $1$  Polya-Gamma latent variable. Alternatively, using calibration parameters $(r,b)$, a  proposal can be simulated from
\be
z &\sim \PG\left ( n r,\theta+b \right),\\
\theta^* &\sim \No \left( \frac{ \sum y_i - r n/2 -z b }{z}, \frac{1}{z}\right),
\ee
and M-H acceptance step as described above using \\ $L_{r,b}(\theta; y) = {\exp(\theta+b)^{\sum y_i}}{\{ 1+\exp(\theta+b)\}^{-nr}}$.
To have a proper $L_{r,b}(\theta;y)$, we further require $r \ge (\sum y_i-1)/n + \epsilon$ with $\epsilon$ a small positive constant. 
%As usual, one would obtain the  Gibbs sampling algorithm \citep{polson2013bayesian} via $(r,b)=(1,0)$. 
 Using Fisher information, the  parameters are adapted initially for $200$
 steps, via 
 
\be
 r & =\frac{\exp(\theta)}{ \{1+\exp(\theta)\} ^2} / \left (   \frac{1}{2 |\theta+b|} \tanh\frac{|\theta+b|}{2} \right) \vee \big ( (\sum y_i-1)/n + \epsilon \big), \\
b & =\log[  \{1+\exp(\theta)\}^{1/r} -1] - \theta.
\ee

\begin{figure}[H]
  {\caption{CDA  maintains high effective sample size, even when scaling
 to massive $n$. Panel(a) shows the performance of DA and CDA when $n$ is scaled up to
 $10^{14}$; Panel(b) shows the performance of CDA and DA, coupled with sub-sampling
 approximation to reduce the number of sampled latent variables. Only accelerating computing time in each iteration (DA-Subsampling) does not solve the scalability issue.      \label{massive_n_sims}}}
  {%
    \subfigure[Effective sample size (with 95\% pointwise credible interval) per $1,000$  steps  with different sample
  size $n$ from $10$ to $10^{14}$, using logistic regression model with intercept only.]{%
      \includegraphics[width=0.45\linewidth]{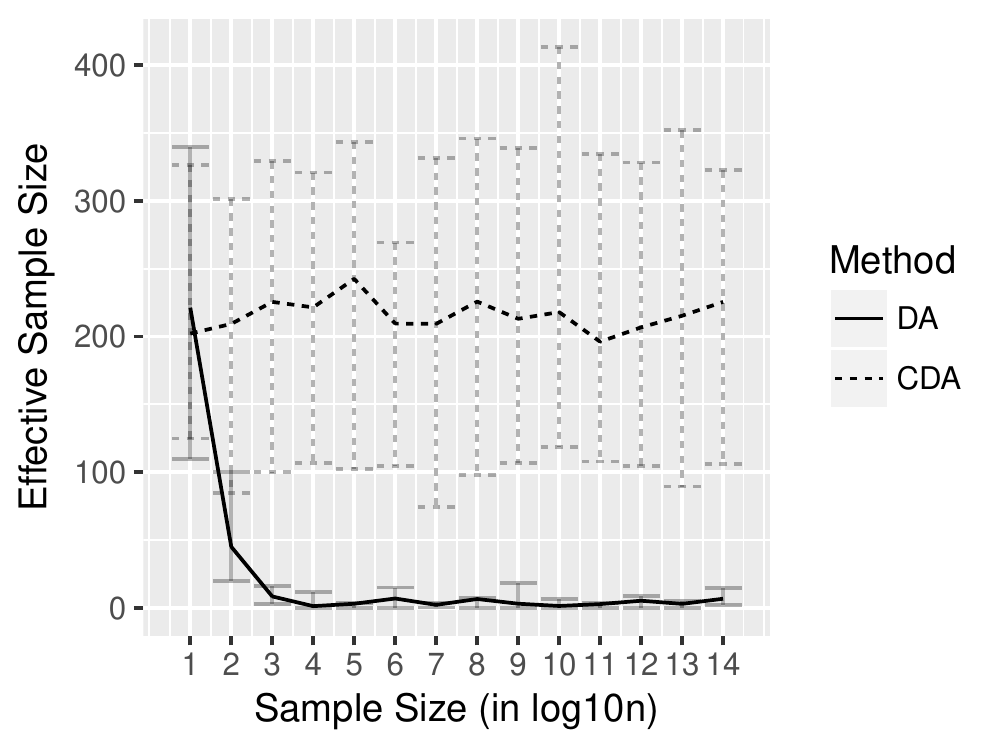}}%
    \qquad
    \subfigure[Effective sample size per (with 95\% pointwise credible interval) $1,000$  steps with different $n$
     from $10^5$ to $10^{8}$, in logistic regression with slope and intercept, using sub-sampling.]{%
      \includegraphics[width=0.55\linewidth]{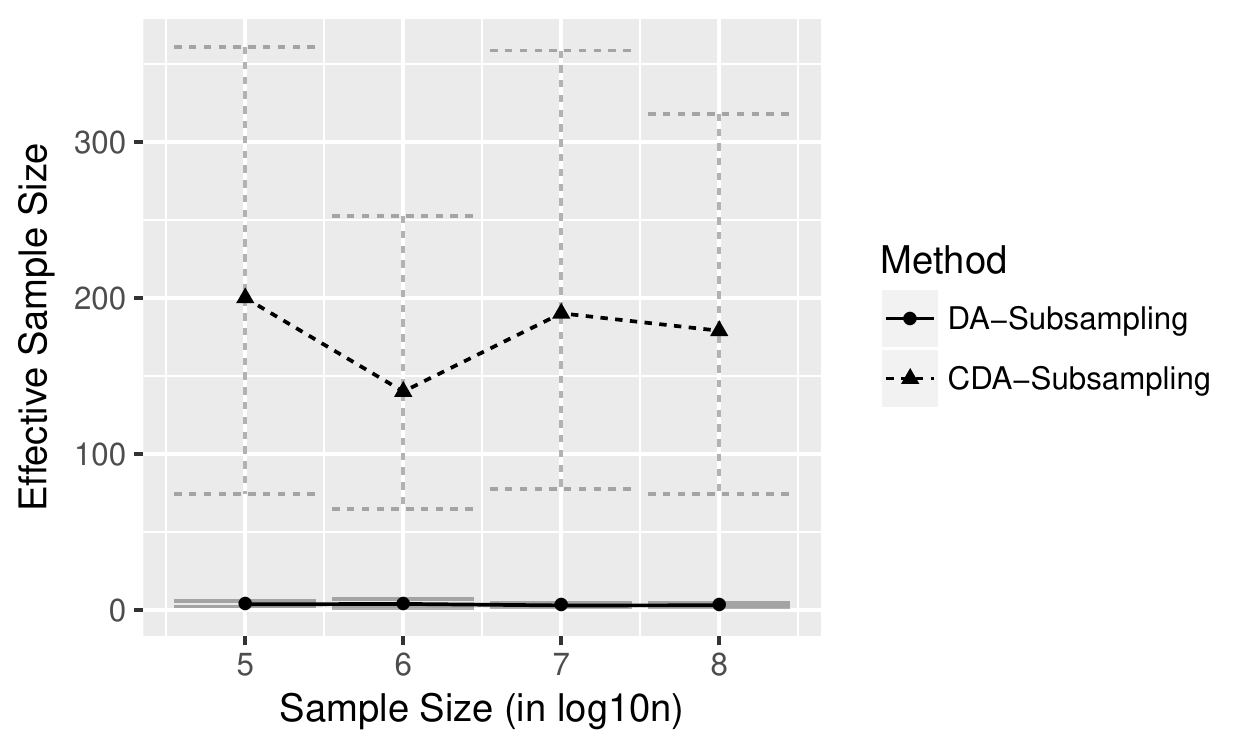}}
  }
\end{figure}

{To obtain enormous data sample size rare event data}, we fixed $\sum
y_i=1$ and increase $n$ from $10^1$ to a massive $10^{14}$. 
Figure~\ref{massive_n_sims}(a) compares the effective sample size per $1,000$
steps using DA and CDA. Surprisingly, the deterioration of DA shows up as early
as $n=10^2$; its slow-down becomes critical at $n=10^4$ with
effective sample size close to $0$. CDA performs exceptionally well, even
at massive $n=10^{14}$ (we stop at $10^{14}$ as $1/n$ reaches the limit of floating
point accuracy).

In more complicated settings, one issue for data augmentation in general is the large number of latent variables to sample in each iteration.  A common strategy is to {avoid sampling latent variables for
every observation by approximating the Markov transition kernel using subsamples} 
\citep{quiroz2016exact,johndrow2015approximations}. {Different from other
example algorithms, this approximation changes the invariant measure. Finding
a suitable sub-sample size while bounding approximation
error requires careful treatment, which is beyond the scope of
this article. Instead, our goal is to show sub-sampling alone does not address the burden of low ESS issue; whereas }{one can trivially couple our proposed CDA
strategy with such subsampling to scale DA-MCMC up to enormous data sample sizes.  We illustrate such 
coupling here.}

We consider the same two-parameter intercept-slope model in logistic regression
as described in the last section, except we now vary data sample size from
$n=10^5$ to $10^8$. We simulate Bernoulli outcome \\$y_i\sim \Bern\big(({1+\exp(-x_i\theta)})^{-1}\big)$ based on $x_1\sim \No(0,1)$ and $\theta=(-\theta_0,1)'$. We vary $\theta_0$ and induce $\sum{y_i}\approx
10$ for each $n$. We utilize the sub-sampled-Polya-Gamma algorithm described by \cite{johndrow2015approximations},
and apply CDA to calibrate the variance discrepancy. Since $y$ is highly imbalanced in
the number of $0$ and $1$s, we apply biased-sampling by including all data
with $y_i=1$, while sub-sampling $1\%$ of data with $y_i=0$. {Existing work on applying biased subsampling in logistic regression mainly aims to obtain point estimates \citep{king2001logistic,wang2017optimal},
in this article we present a simple solution for Bayesian inference.}

{Denoting the set of all data with $y_i=1$ as $V_1$ and a random subset  with $y_i=0$ as $V_0$, it is sensible to keep
the likelihood contribution from $y_i=1$ unchanged, while adjusting the part from $y_i=0$ via a power of a ratio of $({n-|V_1|})/{|V_0|}$, leading to an approximate likelihood
$$L(\theta;y) = \prod_{i\in V_1}\frac{\exp(x_i\theta)}{ 1+\exp(x_i\theta)}  (\prod_{i\in V_0}\frac{1}{ 1+\exp(x_i\theta)}
)^{\frac{n-|V_1|}{|V_0|}}.$$
The number of latent variables is reduced to $|V_0|+|V_1|$; since
$n$ is still large, the slow mixing would remain and calibration is needed.
} The algorithmic details and the calibrated form are presented in the appendix.

Figure~\ref{massive_n_sims}(b) compares the performance of the two approximating
algorithms, one  combining CDA and sub-sampling, and one  using sub-sampling alone. Clearly, only accelerating each step via sub-sampling  does not solve the inefficiency  of very low
effective sample size; while using CDA and  sub-sampling together can produce {excellent computational performance}.

 \section{Co-Browsing Behavior Application}

We apply CDA to an online browsing activity dataset. The dataset contains a two-way  table of visit count by users who browsed one of $96$ client websites of interest, and one of the  $n=59,792$ high-traffic sites during the same browsing session. We refer to visiting more than one site during the same session as co-browsing. For each of the client websites, it is of large commercial interest to find out the high-traffic sites with relatively high co-browsing rates, so that ads can be more effectively placed. For the computational advertising company, it is also useful to understand the co-browsing behavior and predict the traffic pattern of users. We consider two models for these data.

\subsection{Hierarchical Binomial Model for Co-Browsing Rates}

We initially focus on one client website and analyze co-browsing rates with the high-traffic sites. With the total visit count $N_i$ available for the $i$th high-traffic site, the count of co-browsing $y_i$ can be considered as the result of a binomial trial. With $y_i$ extremely small relative to $N_i$ (ratio  $0.00011 \pm  0.00093$), the maximum likelihood estimate $y_i/N_i$ can have poor performance. For example, when $y_i=0$, estimating the rate as exactly $0$ is not ideal. Therefore, it is useful to consider a hierarchical model to allow borrowing of information across high-traffic sites:
\be
y_i \sim \Binom\left(N_i, \frac{\exp(\theta_i)}{1+\exp(\theta_i)}\right), \quad \theta_i\stackrel{iid}{\sim} \No(\theta_0, \sigma^2_0), \quad i=1\ldots n\\
(\theta_0,\sigma^2_0) \sim  \pi(\theta_0,\sigma^2_0) 
\ee
{We choose  weakly informative priors}. Based on expert opinion in quantitative advertising, we use a prior $\theta_0\sim \No(-12,49)$ and uniform prior on $\sigma^2_0$. Similar to the logistic regression, we calibrate the binomial Polya-Gamma augmentation, leading to the proposal likelihood:
\be
L_{r,b}(\theta_i;y_i,N_i, r_i, b_i) = \frac{\exp(\theta_i+b_i)^y_i}{\{ 1+\exp(\theta_i+b_i)\}^{N_ir_i}}
\ee

Conditioned on the latent Polya-Gamma latent variable $z_i$, each proposal $\theta^*_i$ can be sampled from:
\be
z_i &\sim \PG\left( (N_ir_i),\theta_i+b_i \right),\\
\theta_i^* &\sim \No \left( \frac{ y_i - r_i N_i/2 -z_i b_i + \theta_0/\sigma^2_0}{z_i+ 1/\sigma^2_0}, \frac{1}{z_i+ 1/\sigma^2_0}\right),
\ee
and accepted or rejected using an M-H step. We further require $r_i \ge (y_i-1)/N_i + \epsilon$ to have a proper $L_{r,b}(\theta_i;y_i, N_i)$ with $\epsilon$ a small constant. Similar to logistic regression, the auxiliary parameters are chosen as 
\be
r_i & =\frac{\exp(\theta_i)}{ \{1+\exp(\theta_i)\} ^2} / \left (   \frac{1}{2 |\theta_i+b_i|} \tanh\frac{|\theta_i+b_i|}{2} \right) \vee \big ( (y_i-1)/N_i + \epsilon \big),\\
b_i &=\log[  \{1+\exp(\theta_i)\}^{1/r_i} -1] - \theta_i\ee
during adaptation. Since $\theta_i$'s are conditionally independent, the calibrated proposal can be individually accepted with high probability for each $i$. This leads to a high average acceptance of $0.9$, despite the high dimensionality of $59,792$ $\theta_i$'s.
After $\theta_i$'s are updated, other parameters are sampled from 
$\theta_0 \sim \No\big( (n/\sigma^2 +1/49)^{-1} (\sum_i \theta_i /\sigma^2  -12/49 ),  (n /\sigma^2 +1/49)^{-1} \big),$ and $\sigma^2_0 \sim \IG( n/2-1, \sum_i (\theta_i -\theta_0)^2 /2)$.

\begin{figure}[H]
  {\caption{Boxplots of the ACFs show the mixing of the $59,792$ parameters in the hierarchical binomial model, for the original DA\citep{polson2013bayesian}, CDA and HMC. \label{data_binomial}}}
  {%
    \subfigure[ACFs of the rate parameters $\theta_i$ using DA.]{%
      \includegraphics[width=0.32\linewidth]{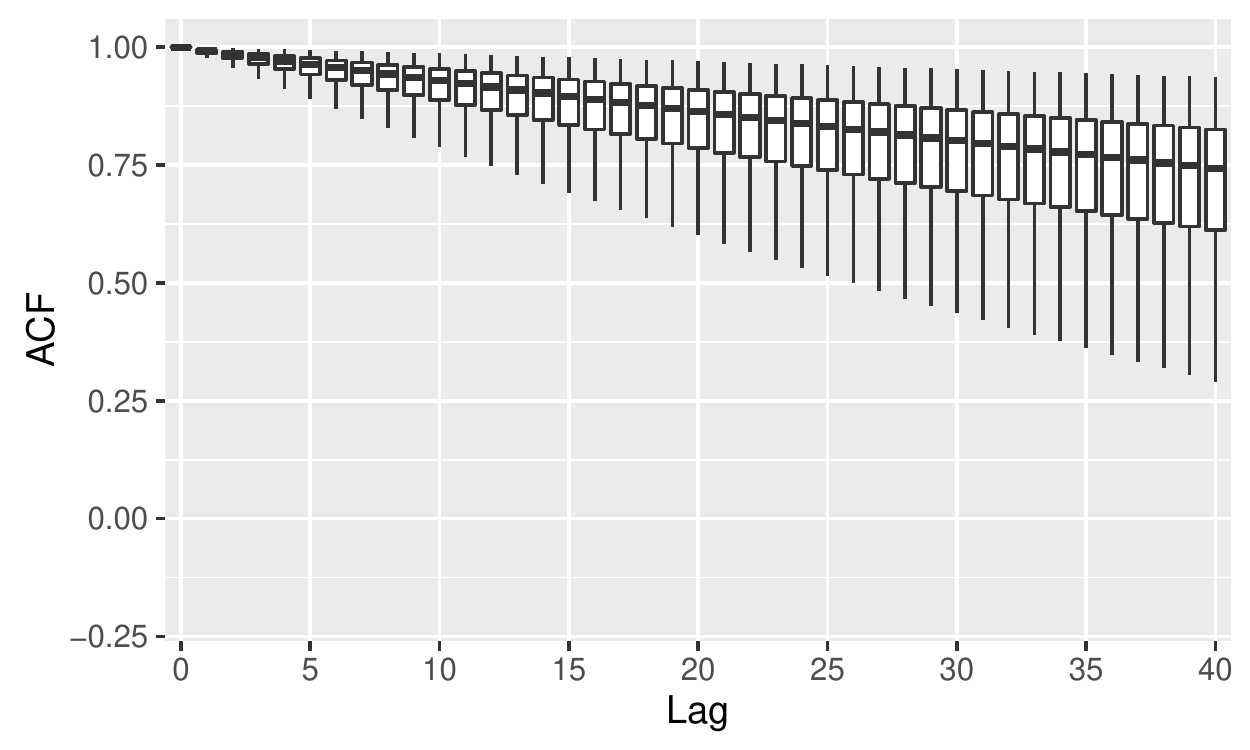}}%
    \subfigure[ACFs of the rate parameters $\theta_i$ using CDA.]{%
      \includegraphics[width=0.32\linewidth]{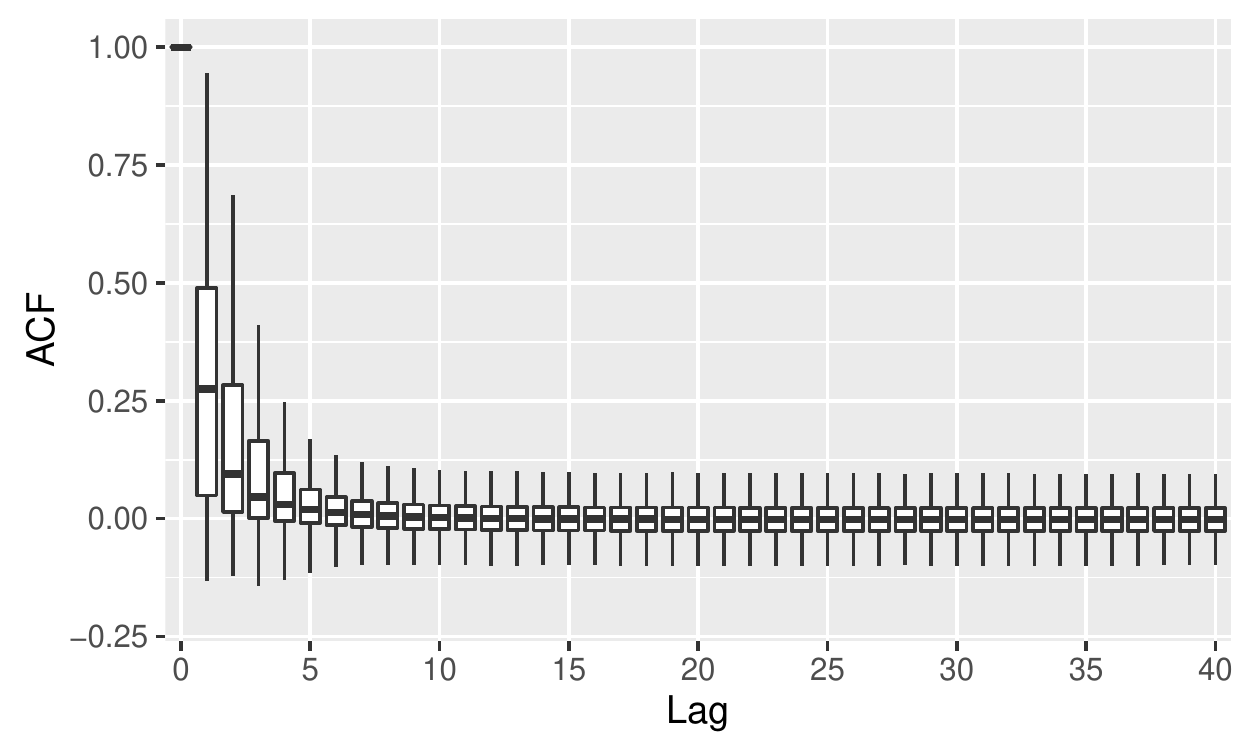}}
    \subfigure[ACFs of the rate parameters $\theta_i$ using HMC.]{%
      \includegraphics[width=0.32\linewidth]{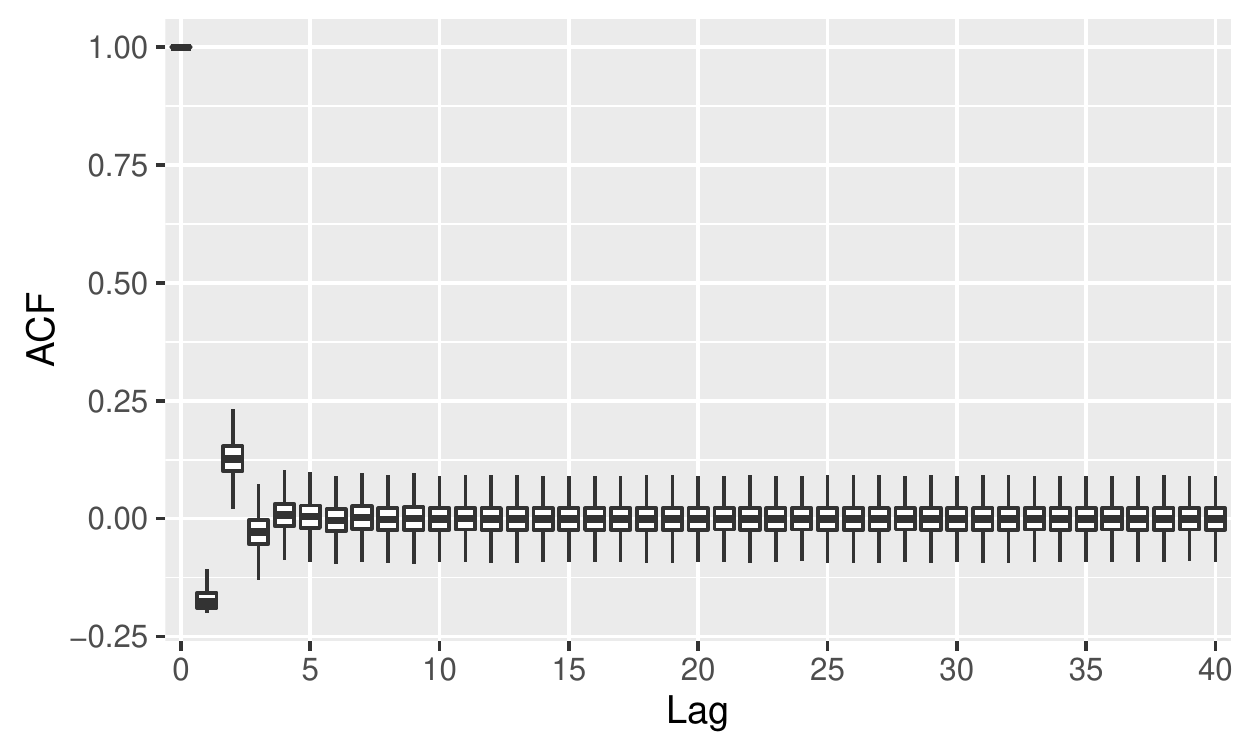}}
  }
\end{figure}

Figure~\ref{data_binomial} shows the boxplots of the ACFs for all $\theta_i$'s. We compare the result with the original DA \citep{polson2013bayesian} and Hamiltonian Monte Carlo (HMC) provided by the \texttt{STAN} software \citep{carpenter2016stan}. We run DA for $100,000$ steps, HMC for $2,000$ steps and CDA for $2,000$ steps, so that they have approximately the same effective sample size (calculated with the \texttt{CODA} package in \texttt{R}). All of the parameters mix poorly in DA; HMC and CDA lead to significant improvement with autocorrelation rapidly decaying to close to zero within $5$ lags.

Shown in Table~\ref{tab:binomial}, CDA and HMC have very close estimates in posterior means and $95\%$ credible intervals for the parameters, while DA has poor estimates due to critically slow mixing. The difference between HMC and CDA is that, although HMC is slightly more efficient in effective sample size per iteration ($T_{eff}/T$) for this model, it is much more computationally intensive and generates many fewer iterations than CDA within the same budget of computing time. As the result, CDA has the most efficient computing time per effective sample.

\begin{table}[H]
\tiny
\centering
\begin{tabular}{|l |r |r| r| r |} 
 \hline
                          & DA & CDA & HMC\\
 [0.5ex]

 $ \sum \theta_i/n$      & -10.03 (-10.16, -9.87)& -12.05 (-12.09, -12.02) &  -12.06 (-12.09, -12.01)\\
 $ \sum \theta_i^2/n$      & 102.25 (98.92, 105.23)& 153.04 (152.06, 154.05) &  153.17 (152.02, 154.29)\\
$\theta_0$          & -10.03 (-10.17, -9.87)& -12.05 (-12.09,-12.01) &  -12.06 (-12.10, -12.01)\\
$\sigma^2$         & 1.60 (1.36, 1.82)&   7.70 (7.49, 7.88)  & 7.71 (7.51, 7.91)\\
$T_{eff} / T$ & 0.0085 (0.0013 0.0188) & 0.5013 (0.1101,1.0084) & 0.8404 (0.5149, 1.2470)\\
Avg Computing Time /  $T$  & 1.2 sec       & 1.2 sec        & 6 sec\\
Avg Computing Time /  $T_{eff}$  & 140.4 sec       & 0.48 sec        & 1.3 sec\\
 \hline
\end{tabular}
\caption{Parameter estimates (with 95\% credible intervals) and computing speed (ratios among computing time, effective sample sizes $T_{eff}$ and total iterations $T$) of the DA, CDA and HMC in hierarchical binomial model. CDA provides parameter estimates as accurate as HMC, and is more computationally efficient than HMC.}
\label{tab:binomial}
\end{table}

\subsection{Poisson Log-Normal Model for Web Traffic Prediction}

The co-browsing on one high-traffic site and one client site is commonly related to the click-through of users from the former to the latter. Therefore, the count of co-browsing is a useful indication of the click-through traffic. For any given client website, predicting the high traffic sites that could generate the most traffic is of high commercial interest. Therefore, we consider a Poisson regression model. We choose the co-browsing count of one client website as the outcome $y_i$ and the log count of the other $95$ websites as the predictors $x_{ij}=\log (x^*_{ij}+1)$ for $i=1,\ldots ,59792$ and $j=1,\ldots ,95$.  A Gaussian random effect is included to account for over-dispersion relative to the Poisson distribution, leading to a Poisson log-normal regression model: 
\be
 y_i \sim \Poi \left( \exp  (\xbeta + \tau_i )\right),  \quad \tau_i\stackrel{iid}{\sim} \No(\tau_0, \nu^2), \quad i=1\ldots n\\
 \beta \sim  \No(0, I \sigma_\beta^2), \quad \tau_0 \sim \No(0,\sigma_\tau^2) \quad \nu^2\sim \pi(\nu^2).
\ee
We assign a weakly informative prior for $\beta$ and $\tau_0$ with $ \sigma_\beta^2=\sigma_\tau^2=100$. For the over-dispersion parameter $\nu^2$, we assign a non-informative uniform prior.

When $\beta$ and $\tau$ are sampled separately, the random effects $\tau = \{\tau_1,\ldots, \tau_n\}$ can cause slow mixing. Instead, we sample $\beta$ and $\tau$ jointly. Using $\tilde X = [ I_n || X ]$ as a $n \times (n+p)$ juxtaposed matrix, and $\eta_i=\xbeta + \tau_i$ for the linear predictor, the model can be viewed as a linear predictor with $n+p$ coefficients, and $\theta= \{\tau, \beta\}'$ can be sampled jointly in a block. The reason for improved mixing with blocked sampling can be found in \cite{liu1994collapsed}.

We now focus on the mixing behavior of data augmentation. We first review data augmentation for the Poisson log-normal model. \cite{zhou2012lognormal} proposed to treat $\Poi(\eta_i)$ as the limit of the negative binomial $\NB\big(\lambda,{\eta_i}/{(\lambda+\eta_i)}\big)$ with $\lambda\rightarrow \infty$, and used moderate $\lambda=1,000$ for approximation. The limit can be simplified as (omitting constant):
\be
L(\eta_i;y_i)=\frac{ \exp(y_i \eta_i \} }{\exp\{\exp(\eta_i)\}} = \lim_{\lambda\rightarrow\infty}\frac{\exp(y_i \eta_i)}{\{1+ \exp(\eta_i )/\lambda\}^{\lambda }}.
\label{eq:pos_approx}
\ee

With finite $\lambda$ approximation, the posterior can be sampled via Polya-Gamma augmented Gibbs sampling:
\be
z_i \mid \eta_i \sim  \PG ( & \lambda, \eta_i -\log \lambda)  \quad i=1\ldots n\\
\theta \mid z, y \sim  \No \bigg(  &  \Big(\tilde X' Z \tilde X+
\begin{bmatrix} 1/\nu^2 \cdot I_n & 0\\ 0 & 1/\sigma^2_{\beta}  \cdot I_p
\end{bmatrix}\Big)^{-1} \\
& \quad \{  \tilde X'  \big ( y - \lambda/2 + Z \log \lambda\big) +   \begin{bmatrix} \tau_0/\nu^2  1_n \\  0_p \end{bmatrix} \} , \\
& \Big(\tilde X' Z \tilde X+  \begin{bmatrix} 1/\nu^2 \cdot I_n & 0\\ 0 & 1/\sigma^2_{\beta}  \cdot I_p \end{bmatrix}\Big)^{-1} \bigg),
\ee
where $Z = \diag\{ z_1, \ldots,  z_n\}$, $1_n = \{1, \ldots 1\}'$ and $0_p = \{0, \ldots 0\}'$.

However, this approximation-based data augmentation is inherently problematic.  For example, setting 
$\lambda = 1,000$ leads to large approximation error.  As in \eqref{eq:pos_approx}, the approximating denominator has $(1+\exp\left(\eta_i)/\lambda\right)^\lambda= \exp \{ \exp(\eta_i) + \bigO(\exp(2\eta_i)/\lambda) \}$; for moderately large $\eta_i \approx 10$, $\lambda$ needs to be at least $10^9$ to make $\exp(2\eta_i)/\lambda$ close to $0$. This large error cannot be corrected with an additional M-H step, since the acceptance rate would be too low. On the other hand, it is not practical to use a large $\lambda$  in a Gibbs sampler, as it would create extremely large $z_i$  (associated with small conditional covariance for $\theta$), resulting in slow mixing.

We use CDA to solve this dilemma. We first choose a very large $\lambda$ ($10^9$) to control the approximation error, then use a small fractional $r_i$ multiplying to $\lambda$ for calibration. This leads to a proposal likelihood similar to the logistic CDA:
\be
L_{r,b}(\xtheta;y_i)=\frac{\exp(\eta_i   -\log \lambda +b_i)^{y_i}}{\{1+ \exp(\eta_i -\log \lambda +b_i)\}^{r_i\lambda  }},
\ee
with $r_i \ge (y_i-1)/\lambda + \epsilon$ for proper likelihood, and proposal update rule:
\be
\small
z_i \sim  \PG ( & r_i\lambda, \eta_i -\log \lambda + b_i)  \quad i=1\ldots n\\
\theta^* \sim  \No \bigg(  &  \Big(\tilde X' Z \tilde X+  \begin{bmatrix}
	1/\nu^2 \cdot I_n & 0\\ 0 & 1/\sigma^2_{\beta}  \cdot I_p
	\end{bmatrix}\Big)^{-1} \\
& \quad \Big\{  \tilde X'  \big( y - r\lambda/2 + Z \log (\lambda -b )\big) +   \begin{bmatrix} \tau_0/\nu^2  1_n \\  0_p \end{bmatrix} \Big\} , \\
& \Big(\tilde X' Z \tilde X+  \begin{bmatrix} 1/\nu^2 \cdot I_n & 0\\ 0 & 1/\sigma^2_{\beta}  \cdot I_p \end{bmatrix}\Big)^{-1} \bigg)
\ee

Letting $\eta_i^* = \tilde X \theta^*$, the proposal is accepted with probability (based on Poisson density and the approximation $L_{r,b}(\xtheta;y_i)$):
\be
1 \wedge \prod_i  \frac{ \exp \{ \exp (\eta_i)\}}{ \exp \{ \exp (\eta_i^*)\}} \frac {{\{1+ \exp(\eta_i^{*}  -\log \lambda +b_i)\}^{r_i\lambda }}}{{\{1+ \exp(\eta_i  -\log \lambda +b_i)\}^{r_i\lambda  }}}.
\ee
 
During the tuning, we set 

\be
r_i & = \tau_i\exp(\eta_i) /  \left( \frac{\lambda } {2|\eta_i + b_i - \log\lambda|}  \tanh\frac{ |\eta_i + b_i - \log\lambda|}{2} \right ) \vee \big( (y_i-1)/\lambda + \epsilon \big), \\
b_i &=\log[ \exp \{ \exp(\eta_i - \log\lambda -\log r_i)   \}-1] -\eta_i + \log\lambda.
\ee
After $\theta$ is updated, the other parameters can be sampled via
\be
\tau_0 &\sim \No\left( (n/ \nu^2 + 1/ \sigma^2_\tau)^{-1} \sum_i
\tau_i/\nu^2 , (n/ \nu^2 + 1/ \sigma^2_\tau)^{-1}  \right)\\
\nu^2 &\sim
\IG ( n/2-1, \sum_i (\tau_i-\tau_0)^2 /2)
\ee

\begin{figure}[H]
  {\caption{CDA significantly improves the mixing of the parameters in the Poisson log-normal. \label{data_poisson}}}
  {%
    \subfigure[Autocorrelation of the parameters from DA.]{%
      \includegraphics[width=0.3\linewidth]{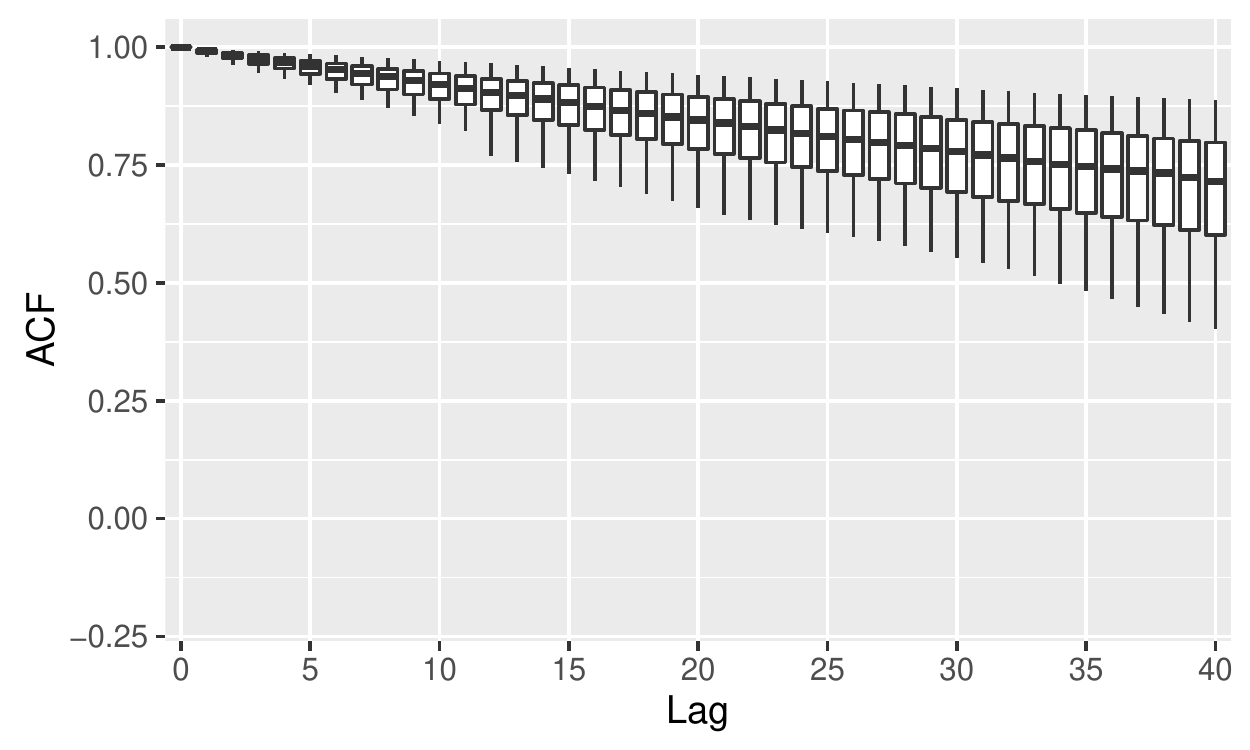}}%
    \subfigure[Autocorrelation of the parameters from CDA.]{%
      \includegraphics[width=0.3\linewidth]{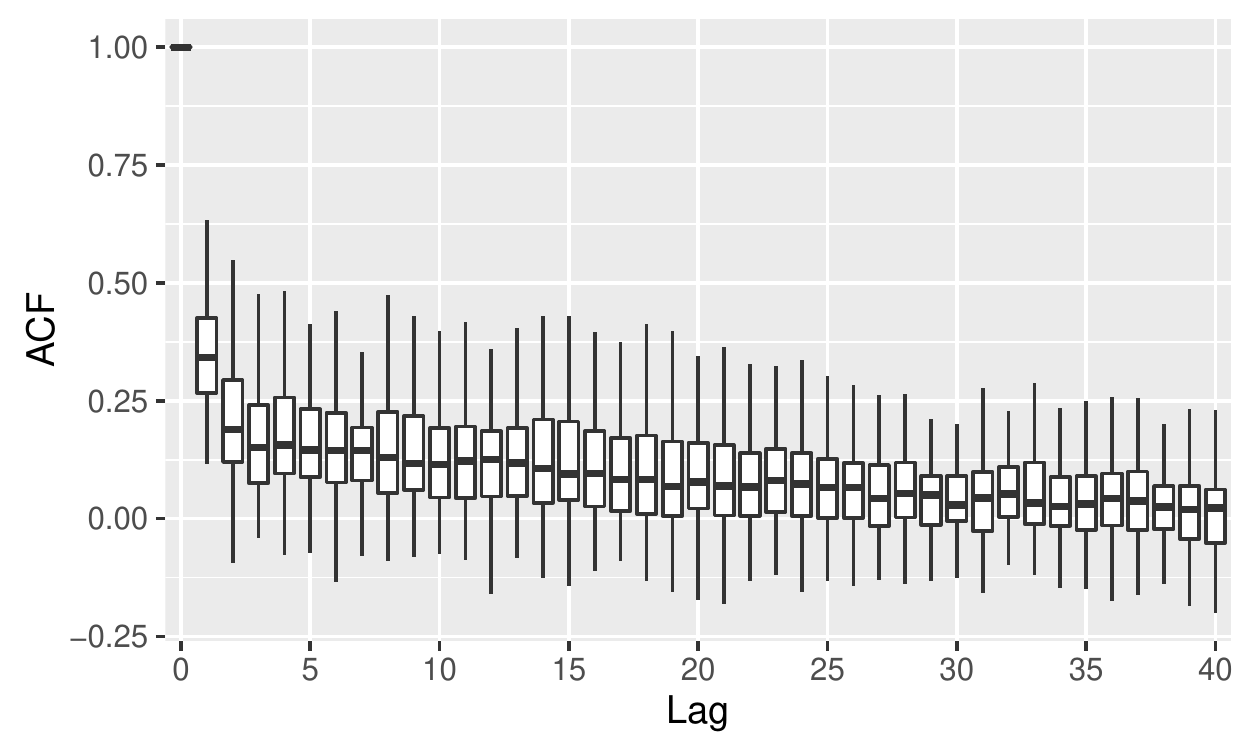}}
    \subfigure[Autocorrelation of the parameters from HMC.]{%
      \includegraphics[width=0.3\linewidth]{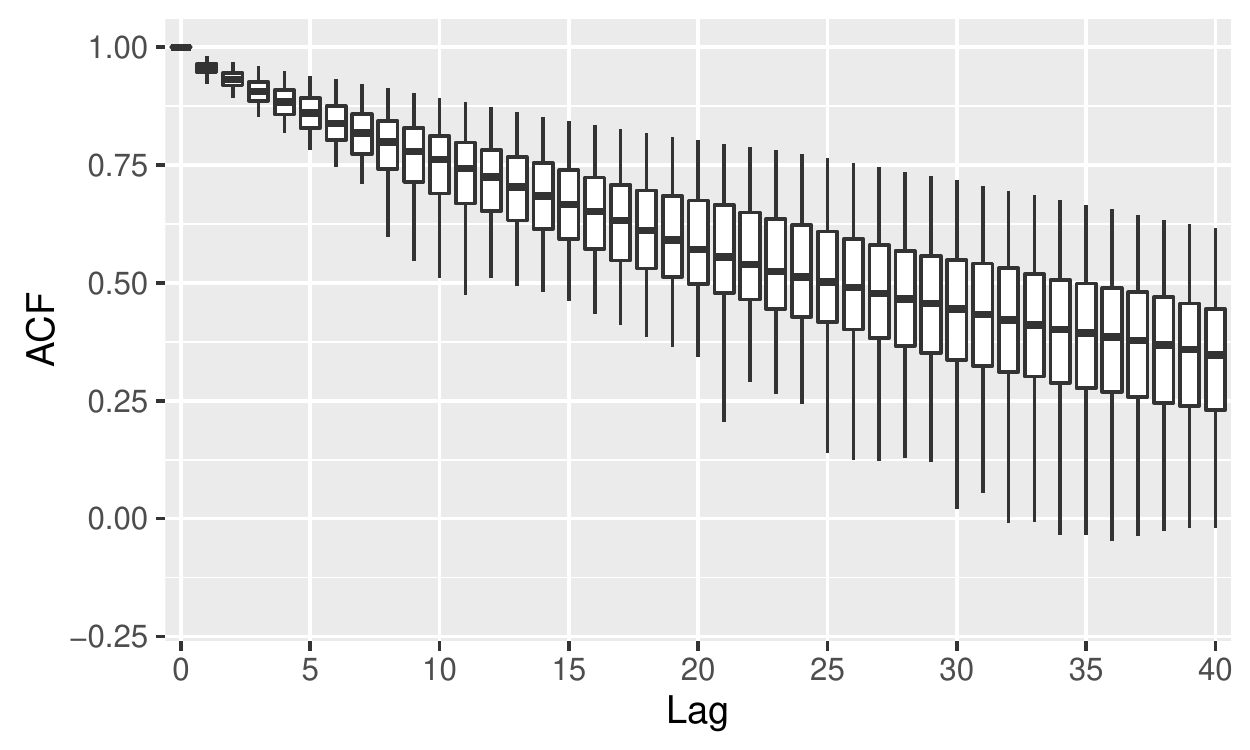}}
  }
\end{figure}

We ran the basic DA with $\lambda=1,000$ approximation, CDA with $\lambda=10^9$ and HMC. We ran DA for $200,000$ steps, CDA for $2,000$ steps and HMC for $20,000$ steps so that they have approximately the same effective sample size. For CDA, we used the first $1,000$ steps for adapting $r$ and $b$. Figure~\ref{data_poisson} shows the mixing of DA, CDA and HMC. Even with small $\lambda = 1,000$ in DA, all of the parameters mix poorly; HMC seemed to be affected by the presence of random effects, and most of parameters remain highly correlated within $40$ lags; CDA substantially improves the mixing. Table~\ref{table:Poisson} compares all three algorithms. CDA has the most efficient computing time per effective sample, and is about $30-300$ times more efficient than the other two algorithms.

To evaluate the prediction performance, we use another co-browsing count table for the same high traffic and client sites, collected during a different time period. We use the high traffic co-browsing count $x_{ij}^{\dagger*}$ and their log transform 
$x^\dagger_{ij} = \log(   x_{ij}^{\dagger*} +1 )$ for the  $j = 1,\ldots,95$ clients to predict the count for the client of interest $y_i^\dagger$, over the high traffic site $i=1,\ldots, 59792$. We carry out prediction using $\hat y_i^\dagger= \bb E_{ \beta, \tau \mid y,x}   y_{i}^\dagger =\bb E_{ \beta, \tau \mid y,x}\exp(  x_{i}^\dagger\beta + \tau_i)$ on the client site. The expectation is taken over posterior sample $\beta, \tau \mid y,x$ with training set $\{y,x\}$ discussed above. Cross-validation root-mean-squared error $\big(\sum_i(\hat y_i^\dagger - y_i^\dagger)^2/n\big)^{1/2}$ between the prediction and actual count $ y_i^\dagger$'s is computed. Shown in Table~\ref{table:Poisson}, slow mixing in DA and HMC cause poor estimation of the parameters and high prediction error, while CDA has significantly lower error. 

\begin{table}[H]
\tiny
\centering
\begin{tabular}{|l |r |r| r| r |} 
 \hline
                          & DA & CDA & HMC\\
 [0.5ex]
$\sum \beta_j / 95$         & 0.072 (0.071, 0.075)&  -0.041 (-0.042, -0.038)  & -0.010 (-0.042, -0.037) \\
$\sum \beta_j^2 / 95$         & 0.0034 (0.0033, 0.0035)&  0.231 (0.219 0.244)  & 0.232 (0.216 0.244)   \\
$\sum\tau_i/n$         & -0.405 (-0.642, -0.155)&  -1.292 (-2.351, -0.446)  &  -1.297 (-2.354, -0.451)  \\
$\sum\tau_i^2/n$         & 1.126 (0.968, 1.339)&  3.608 (0.696, 7.928)  & 3.589 (0.678, 8.011)  \\
% RMSE                              & 30.83        & 4.03          & 7.38\\
Prediction RMSE                           & 33.21        & 8.52          & 13.18\\
$T_{eff} / T$ & 0.0037 (0.0011 0.0096) & 0.3348 (0.0279, 0.699) &  0.0173 (0.0065, 0.0655) \\
Avg Computing Time /  $T$  & 1.3 sec       & 1.3 sec        & 56 sec\\
Avg Computing Time /  $T_{eff}$  & 346.4 sec       & 11.5 sec        & 3240.6 sec\\
 \hline
\end{tabular}
\caption{Parameter estimates, prediction error and computing speed of the DA, CDA and HMC in Poisson regression model.}
\label{table:Poisson}
\end{table}

\section{Discussion}
Data augmentation (DA) is a technique routinely used to enable implementation of simple Gibbs samplers, avoiding the need for expensive and complex tuning of Metropolis-Hastings algorithms.
Despite the convenience, DA can slow down mixing when the conditional posterior variance given the augmented data is substantially smaller than the marginal variance.  When the data sample size is massive, this problem arises when the rates of convergence of the augmented and marginal posterior differ, leading to 
critical mixing problems.  There is a very rich literature on strategies for improving mixing rates of Gibbs samplers, with centered or non-centered re-parameterizations \citep{papaspiliopoulos2007general} and parameter-expansion \citep{liu1999parameter} leading to some improvements.  However, existing approaches {do not} solve large sample mixing problems in not addressing the fundamental rate
mismatch issue.

To tackle this problem, we propose to calibrate data augmentation and use a parameter to directly adjust the conditional variance (which is associated with step size).  CDA adds a little cost due to the likelihood evaluation, which is often negligible as compared to the random number generation. In this article, we demonstrate that calibration is generally applicable when $\theta \mid z$ belongs to the location-scale family. We expect it to be extensible to any conditional distribution with a variance or scale.

As both CDA and HMC involve M-H steps, we draw some further comparison between the two. Both methods rely on finding a good proposal by searching a region far from the current state. One key difference lies in the computing efficiency. Although HMC is more generally applicable beyond data augmentation, it is computationally intensive since Hamiltonian dynamics often requires multiple numeric steps. CDA only requires one step of calibrated Gibbs sampling, which is often much more efficient leveraging on existing data augmentation algorithms.

%\bigskip
%\begin{center}
%{\large\bf SUPPLEMENTARY MATERIAL}
%\end{center}

\appendix
\section{Appendix}

\subsection{Proof of Remark \ref{rem:accrat}}
\begin{proof}
Since $Q_{r,b}(\theta;\theta')$ is the $\theta$ marginal of a Gibbs transition kernel, and Gibbs is reversible on its margins, we have
\be
Q(\theta;\theta') \Pi_{r,b}(\theta) = Q(\theta';\theta) \Pi_{r,b}(\theta),
\ee 
and so
\be
\frac{L(\theta';y) \Pi^0(\theta') Q(\theta;\theta')}{L(\theta;y) \Pi^0(\theta) Q(\theta';\theta)} &= \frac{L(\theta';y) \Pi^0(\theta') L_{r,b}(\theta;y) \Pi^0(\theta) }{L(\theta;y) \Pi^0(\theta) L_{r,b}(\theta';y) \Pi^0(\theta')} \\
&= \frac{L(\theta';y)L_{r,b}(\theta;y)}{L(\theta;y) L_{r,b}(\theta';y)}.
\ee
\end{proof}

\subsection{Proof of Remark \ref{rem:ergodic}}
\begin{proof}
For any $r,b$, the conditionals $\Pi_{r,b}(z \mid \theta)$ and $\Pi_{r,b}(\theta \mid z)$ are well-defined for all $z \in \mc Z, \theta \in \Theta$, and therefore the Gibbs transition kernel $K_{r,b}((\theta,z);\cdot)$ and corresponding marginal kernels $Q_{r,b}(\theta;\cdot)$ are well-defined. Moreover, for any $(z,\theta) \in \mc Z \times \Theta$, we have $\bb P[(\theta',z') \in A \mid (\theta,z)] > 0$ by assumption. Thus $K_{r,b}$ is aperiodic and $\Pi_{r,b}$-irreducible (see the discussion following Corollary 1 in \cite{roberts1994simple}).

$Q_{r,b}(\theta';\theta)$ is aperiodic and $\Pi_{r,b}(\theta)$-irreducible, since it is the $\theta$ marginal transition kernel induced by $K_{r,b}((\theta,z);\cdot)$. Thus, it is also $\Pi(\theta)$-irreducible so long as $\Pi \gg \Pi_{r,b}$, where for two measure $\mu,\nu$, $\mu \gg \nu$ indicates absolute continuity. Since $\Pi, \Pi_{r,b}$ have densities with respect to Lebesgue measure, $\Pi_{r,b}$-irreducibility implies $\Pi$ irreducibility. Moreover, $Q(\theta;\theta') > 0$ for all $\theta \in \Theta$. Thus, by \cite[Theorem 3]{roberts1994simple}, CDA M-H is $\Pi$-irreducible and aperiodic. 
\end{proof}

\begin{figure}[H]
  {\caption { Diagnostics plot of the adaptation for probit regression. \label{probitAdaptDiag}}}
  {%
    \subfigure[Posterior sample of $x_i\beta$ and its corresponding transform
$({x_i\beta+b_i})/{\sqrt{r_i}}$  in probit CDA.]{%
      \includegraphics[width=0.45\linewidth]{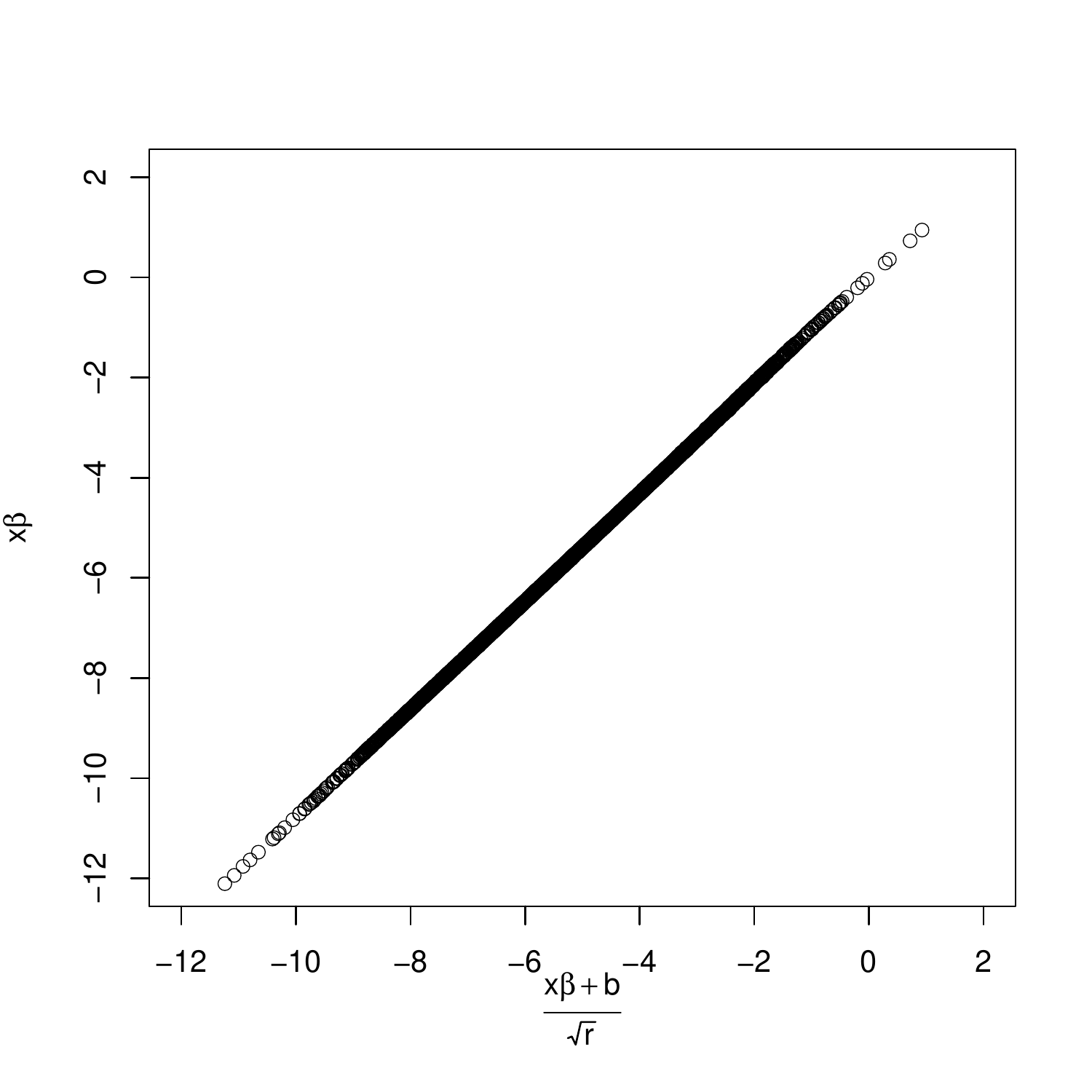}}%
    \qquad
    \subfigure[Trace of the adaptation of r, shown on log scale.]{%
      \includegraphics[width=0.45\linewidth]{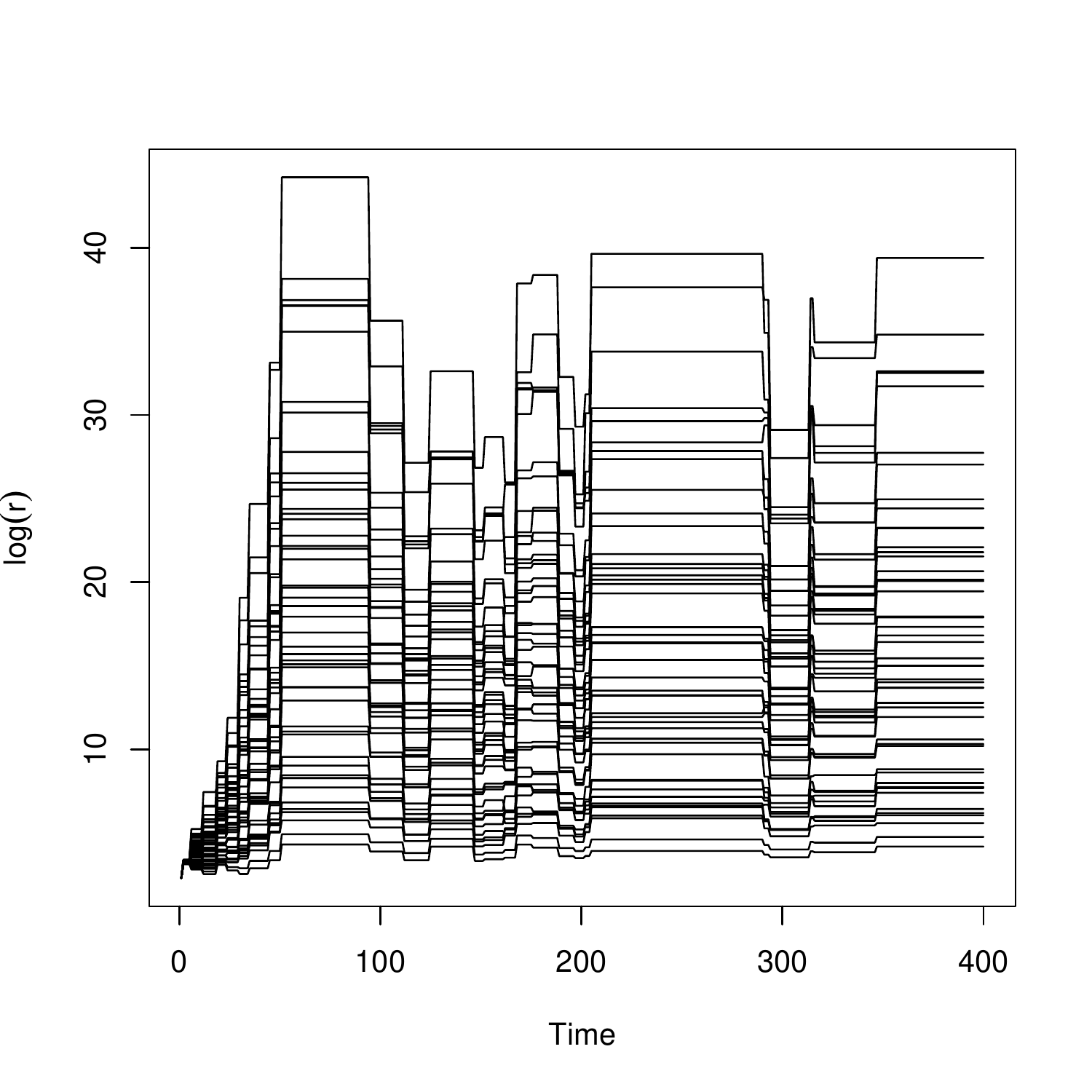}}
  }
\end{figure}

{
\subsection{Diagnostics of Adaptation}
}
We adapt the tuning parameters $(r,b)$ by locally minimizing difference between two Fisher information matrices and
 optimizing acceptance rate near 
 $\hat\theta_{MAP}$ during adaptation. Updating $b$  as a function
 of $\theta$ yields $L_{r,b}(\theta;y)=L(\theta;y)$ exactly;  after adaptation,
for  fixed
 $(r,b)$, $L_{r,b}(\theta;y)$ is close to $L(\theta;y)$ for $\theta$ in the neighborhood
 around $\hat\theta_{MAP}$. To show this empirically, consider the
 probit Bernoulli regression example. As the $L(\theta;y)$ and $L_{r,b}(\theta;y)$
 are parameterized by $\Phi(x_i\beta)$ and $\Phi(({x_i\beta+b_i})/{\sqrt{r_i}})$, Figure~\ref{probitAdaptDiag}(a) compares the posterior values of $x_i\beta$ against
$({x_i\beta+b_i})/{\sqrt{r_i}}$. Clearly, the two are very close with a
RMSE of $0.23$.
Figure~\ref{probitAdaptDiag}(b) shows the trace of the adaptation of $r$
during
the initial $400$ iterations;
$r$ quickly rises from $1$ to the roughly  appropriate scale during the
initial $50$ steps. The values of $(r,b)$ are fixed afterwards to ensure
ergodicity.

{
\subsection{Calibrated  Polya-Gamma Algorithm with Sub-sampling}
}
Adapting based on \cite{johndrow2015approximations}, we first randomly sample
a subset of indices $V$ of size $|V|$. This algorithm  generates proposals from

\be
V & = V_1\cup V_0,\quad V_1= \{i\in\{1,\ldots,n\}: y_i=1\}, \\ \quad V_0 & \sim \text{Subset}(|V|,\{i\in\{1,\ldots,n\}: y_i=0\})
\\ z_i &\sim {\PG}(k_{i}r_i, |\xtheta+b_i|) \quad i\in V,\\
\theta^* &\sim \No \left(  (X_V' Z_{V} X_V)^{-1}  X_V'  (y_V -k_{V}r_V/2- Z_Vb_V) ,  (X_V' Z_V X_V)^{-1}  \right),
\ee
where subscript $._V$ indicates the sub-matrix or sub-vector corresponding
to the sub-sample; $k_i=1$ if $y_i=1$, and
$k_i=({n-|V_1|})/{|V_0|}$. We accept $\theta^*$ in M-H step using calibrated
likelihood 
$$L_{r,b}(\theta;y) = \prod_{i\in V_1}\frac{\exp(x_i\theta+b_i)}{\{ 1+\exp(x_i\theta+b_i)\}^{r_i}}  (\prod_{i\in V_0}\frac{1}{\{ 1+\exp(x_i\theta+b_i)\}^{r_i}}
)^{\frac{n-|V_1|}{|V_0|}}
,$$
with target approximate likelihood  $L_{1,0}(\theta;y)$.
Using Fisher information, the  parameters are adapted initially for $200$
 steps, via 
 \be r_{i} & =\frac{\exp(x_i\theta)}{ \{1+\exp(x_i\theta)\} ^2} / \left (   \frac{1}{2 |x_i\theta+b_{i}|} \tanh\frac{|x_i\theta+b_{i}|}{2} \right) \vee \big ( ( y_i-1)/k_{i} + \epsilon \big)\\
b_{i} & =\log[  \{1+\exp(x_i\theta+b_{i})\}^{1/r_{i}} -1] - x_i\theta.\ee

\bibliography{reference}

\begin{thebibliography}{}

\bibitem[\protect\citeauthoryear{Albert and Chib}{Albert and
  Chib}{1993}]{albert1993bayesian}
Albert, J.~H. and S.~Chib (1993).
\newblock {Bayesian analysis of binary and polychotomous response data}.
\newblock {\em Journal of the American Statistical Association\/}~{\em
  88\/}(422), 669--679.

\bibitem[\protect\citeauthoryear{Atkinson and Mitchell}{Atkinson and
  Mitchell}{1981}]{atkinson1981rao}
Atkinson, C. and A.~F. Mitchell (1981).
\newblock {Rao's distance measure}.
\newblock {\em Sankhy{\=a}: The Indian Journal of Statistics, Series A\/},
  345--365.

\bibitem[\protect\citeauthoryear{Carpenter, Gelman, Hoffman, Lee, Goodrich,
  Betancourt, Brubaker, Guo, Li, and Riddell}{Carpenter
  et~al.}{2016}]{carpenter2016stan}
Carpenter, B., A.~Gelman, M.~Hoffman, D.~Lee, B.~Goodrich, M.~Betancourt, M.~A.
  Brubaker, J.~Guo, P.~Li, and A.~Riddell (2016).
\newblock {STAN: a probabilistic programming language}.
\newblock {\em Journal of Statistical Software\/}.

\bibitem[\protect\citeauthoryear{Conrad, Marzouk, Pillai, and Smith}{Conrad
  et~al.}{2015}]{conrad2015accelerating}
Conrad, P.~R., Y.~M. Marzouk, N.~S. Pillai, and A.~Smith (2015).
\newblock {Accelerating asymptotically exact MCMC for computationally intensive
  models via local approximations}.
\newblock {\em Journal of the American Statistical Association\/}~(1591--1607).

\bibitem[\protect\citeauthoryear{Efron and Hinkley}{Efron and
  Hinkley}{1978}]{efron1978assessing}
Efron, B. and D.~V. Hinkley (1978).
\newblock {Assessing the accuracy of the maximum likelihood estimator: observed
  versus expected Fisher information}.
\newblock {\em Biometrika\/}~{\em 65\/}(3), 457--483.

\bibitem[\protect\citeauthoryear{Johndrow, Mattingly, Mukherjee, and
  Dunson}{Johndrow et~al.}{2017}]{johndrow2015approximations}
Johndrow, J.~E., J.~C. Mattingly, S.~Mukherjee, and D.~B. Dunson (2017).
\newblock {Optimal approximating Markov chains for Bayesian inference}.
\newblock {\em arXiv preprint arXiv:1508.03387\/}.

\bibitem[\protect\citeauthoryear{Johndrow, Smith, Pillai, and Dunson}{Johndrow
  et~al.}{2016}]{johndrow2016inefficiency}
Johndrow, J.~E., A.~Smith, N.~Pillai, and D.~B. Dunson (2016).
\newblock {Inefficiency of data augmentation for large sample imbalanced data}.
\newblock {\em arXiv preprint arXiv:1605.05798\/}.

\bibitem[\protect\citeauthoryear{King and Zeng}{King and
  Zeng}{2001}]{king2001logistic}
King, G. and L.~Zeng (2001).
\newblock {Logistic regression in rare events data}.
\newblock {\em Political Analysis\/}~{\em 9\/}(2), 137--163.

\bibitem[\protect\citeauthoryear{Liu}{Liu}{1994a}]{liu1994collapsed}
Liu, J.~S. (1994a).
\newblock {The collapsed Gibbs sampler in Bayesian computations with
  applications to a gene regulation problem}.
\newblock {\em Journal of the American Statistical Association\/}~{\em
  89\/}(427), 958--966.

\bibitem[\protect\citeauthoryear{Liu}{Liu}{1994b}]{liu1994fraction}
Liu, J.~S. (1994b).
\newblock {The fraction of missing information and convergence rate for data
  augmentation}.
\newblock {\em Computing Science and Statistics\/}, 490--490.

\bibitem[\protect\citeauthoryear{Liu and Wu}{Liu and
  Wu}{1999}]{liu1999parameter}
Liu, J.~S. and Y.~N. Wu (1999).
\newblock {Parameter expansion for data augmentation}.
\newblock {\em Journal of the American Statistical Association\/}~{\em
  94\/}(448), 1264--1274.

\bibitem[\protect\citeauthoryear{Maclaurin and Adams}{Maclaurin and
  Adams}{2015}]{maclaurin2014firefly}
Maclaurin, D. and R.~P. Adams (2015).
\newblock {Firefly Monte Carlo: exact MCMC with subsets of data}.
\newblock In {\em Proceedings of the Twenty-Fourth International Joint
  Conference on Artificial Intelligence}, pp.\  543--552.

\bibitem[\protect\citeauthoryear{Meng and Van~Dyk}{Meng and
  Van~Dyk}{1999}]{meng1999seeking}
Meng, X.-L. and D.~A. Van~Dyk (1999).
\newblock {Seeking efficient data augmentation schemes via conditional and
  marginal augmentation}.
\newblock {\em Biometrika\/}~{\em 86\/}(2), 301--320.

\bibitem[\protect\citeauthoryear{Minsker, Srivastava, Lin, and Dunson}{Minsker
  et~al.}{2014}]{minsker2014robust}
Minsker, S., S.~Srivastava, L.~Lin, and D.~B. Dunson (2014).
\newblock {Robust and scalable Bayes via a median of subset posterior
  measures}.
\newblock {\em arXiv preprint arXiv:1403.2660\/}.

\bibitem[\protect\citeauthoryear{Ngai, Hu, Wong, Chen, and Sun}{Ngai
  et~al.}{2011}]{ngai2011application}
Ngai, E., Y.~Hu, Y.~Wong, Y.~Chen, and X.~Sun (2011).
\newblock {The application of data mining techniques in financial fraud
  detection: a classification framework and an academic review of literature}.
\newblock {\em Decision Support Systems\/}~{\em 50\/}(3), 559--569.

\bibitem[\protect\citeauthoryear{Papaspiliopoulos, Roberts, and
  Sk{\"o}ld}{Papaspiliopoulos et~al.}{2007}]{papaspiliopoulos2007general}
Papaspiliopoulos, O., G.~O. Roberts, and M.~Sk{\"o}ld (2007).
\newblock {A general framework for the parametrization of hierarchical models}.
\newblock {\em Statistical Science\/}, 59--73.

\bibitem[\protect\citeauthoryear{Polson, Scott, and Windle}{Polson
  et~al.}{2013}]{polson2013bayesian}
Polson, N.~G., J.~G. Scott, and J.~Windle (2013).
\newblock {Bayesian inference for logistic models using P{\'o}lya--Gamma latent
  variables}.
\newblock {\em Journal of the American Statistical Association\/}~{\em
  108\/}(504), 1339--1349.

\bibitem[\protect\citeauthoryear{Quiroz, Villani, and Kohn}{Quiroz
  et~al.}{2016}]{quiroz2016exact}
Quiroz, M., M.~Villani, and R.~Kohn (2016).
\newblock {Exact subsampling MCMC}.
\newblock {\em arXiv preprint arXiv:1603.08232\/}.

\bibitem[\protect\citeauthoryear{Roberts and Rosenthal}{Roberts and
  Rosenthal}{2007}]{roberts2007coupling}
Roberts, G.~O. and J.~S. Rosenthal (2007).
\newblock {Coupling and ergodicity of adaptive Markov chain Monte Carlo
  algorithms}.
\newblock {\em Journal of Applied Probability\/}, 458--475.

\bibitem[\protect\citeauthoryear{Roberts and Smith}{Roberts and
  Smith}{1994}]{roberts1994simple}
Roberts, G.~O. and A.~F. Smith (1994).
\newblock {Simple conditions for the convergence of the Gibbs sampler and
  Metropolis-Hastings algorithms}.
\newblock {\em Stochastic Processes and Their Applications\/}~{\em 49\/}(2),
  207--216.

\bibitem[\protect\citeauthoryear{Rubin}{Rubin}{2004}]{rubin2004multiple}
Rubin, D.~B. (2004).
\newblock {\em {Multiple imputation for nonresponse in surveys}}, Volume~81.
\newblock John Wiley \& Sons.

\bibitem[\protect\citeauthoryear{Srivastava, Cevher, Tran-Dinh, and
  Dunson}{Srivastava et~al.}{2015}]{srivastava2015wasp}
Srivastava, S., V.~Cevher, Q.~Tran-Dinh, and D.~B. Dunson (2015).
\newblock {WASP: scalable Bayes via Barycenters of subset posteriors}.
\newblock In {\em Proceedings of the Eighteenth International Conference on
  Artificial Intelligence and Statistics}, pp.\  912--920.

\bibitem[\protect\citeauthoryear{Tanner and Wong}{Tanner and
  Wong}{1987}]{tanner1987calculation}
Tanner, M.~A. and W.~H. Wong (1987).
\newblock {The calculation of posterior distributions by data augmentation}.
\newblock {\em Journal of the American Statistical Association\/}~{\em
  82\/}(398), 528--540.

\bibitem[\protect\citeauthoryear{Tran, Pitt, and Kohn}{Tran
  et~al.}{2016}]{tran2016adaptive}
Tran, M.-N., M.~K. Pitt, and R.~Kohn (2016).
\newblock {Adaptive Metropolis--Hastings sampling using reversible dependent
  mixture proposals}.
\newblock {\em Statistics and Computing\/}~{\em 26\/}(1-2), 361--381.

\bibitem[\protect\citeauthoryear{Wakefield}{Wakefield}{2007}]{wakefield2007disease}
Wakefield, J. (2007).
\newblock {Disease mapping and spatial regression with count data}.
\newblock {\em Biostatistics\/}~{\em 8\/}(2), 158--183.

\bibitem[\protect\citeauthoryear{Wang, Zhu, and Ma}{Wang
  et~al.}{2017}]{wang2017optimal}
Wang, H., R.~Zhu, and P.~Ma (2017).
\newblock {Optimal subsampling for large sample logistic regression}.
\newblock {\em Journal of the American Statistical Association\/}~(in press).

\bibitem[\protect\citeauthoryear{Wang, Li, Cui, Zhang, and Mao}{Wang
  et~al.}{2010}]{wang2010click}
Wang, X., W.~Li, Y.~Cui, R.~Zhang, and J.~Mao (2010).
\newblock {Click-through rate estimation for rare events in online
  advertising}.
\newblock {\em Online Multimedia Advertising: Techniques and Technologies\/},
  1--12.

\bibitem[\protect\citeauthoryear{Zhou, Li, Dunson, and Carin}{Zhou
  et~al.}{2012}]{zhou2012lognormal}
Zhou, M., L.~Li, D.~B. Dunson, and L.~Carin (2012).
\newblock {Lognormal and Gamma mixed negative Binomial regression}.
\newblock In {\em Proceedings of the International Conference on Machine
  Learning}, Volume 2012, pp.\  1343.

\end{thebibliography}
\bibliographystyle{chicago}
\end{document}